\documentclass[dvipsnames,aps,prl,nobibnotes,twocolumn,superscriptaddress,nolongbibliography]{revtex4-2}
\usepackage[T1]{fontenc}
\usepackage[utf8]{inputenc}

\usepackage[unicode=true,pdfusetitle,
 bookmarks=true,bookmarksnumbered=false,bookmarksopen=false,
 breaklinks=false,pdfborder={0 0 0},pdfborderstyle={},backref=false,colorlinks=true]
 {hyperref}
 \hypersetup{
 citecolor=blue,
 urlcolor=blue,
 linkcolor=blue}
\usepackage[table,svgnames]{xcolor}
\usepackage{array}
\usepackage{units}
\usepackage{amsmath}
\usepackage{amssymb}
\usepackage{graphicx}
\usepackage{stackrel}
\usepackage{graphicx}
\usepackage{babel}
\usepackage[final]{pdfpages}
\usepackage{mathtools}
\newcommand{\new}[1]{{\color{black} {#1}}}

\makeatletter
\AtBeginDocument{\let\LS@rot\@undefined}
\makeatother

\begin{document}

\author{Sandro Meloni}
\email[]{sandro@ifisc.uib-csic.es}

\affiliation{Institute for Cross-Disciplinary Physics and Complex Systems (IFISC), CSIC-UIB, 07122 - Palma de Mallorca, Spain}
\affiliation{``Enrico Fermi'' Research Center (CREF), Via Panisperna 89A, 00184 - Rome, Italy}
\affiliation{Institute for Applied Mathematics ``Mauro Picone'' (IAC) CNR, Via dei Taurini 19, 00185 - Rome, Italy}
\author{Andrea Gabrielli}
\affiliation{``Enrico Fermi'' Research Center (CREF), Via Panisperna 89A, 00184 - Rome, Italy}
\affiliation{Dipartimento di Ingegneria Civile, Informatica e delle Tecnologie Aeronautiche, Universit\`a degli Studi ``Roma Tre'', Via Vito Volterra 62, 00146 - Rome, Italy.}
\affiliation{Istituto dei Sistemi Complessi (ISC) - CNR, Rome, Italy.}
\author{Pablo Villegas}
\email[]{pablo.villegas@cref.it}
\affiliation{``Enrico Fermi'' Research Center (CREF), Via Panisperna 89A, 00184 - Rome, Italy}
\affiliation{Instituto Carlos I de F\'isica Te\'orica y Computacional, Univ. de Granada, E-18071, Granada, Spain.}

\begin{abstract}
    Higher-order interactions have recently emerged as a promising framework for describing new dynamical phenomena in heterogeneous contagion processes. However, a fundamental open question is how to understand their contribution from the perspective of the physics of critical phenomena. Based on a mesoscopic field-theoretic Langevin description, we show that: (i) pairwise mechanisms such as facilitation or thresholding are formally equivalent to higher-order ones, (ii) pairwise interactions at coarse-grained scales govern the \new{simplicial} contact process and, (iii) the interplay between noise and topology is determined by the network spectral dimension. In short, we demonstrate that classical field theories, rooted in model symmetries and/or network dimensionality, capture the nature of the phase transition in real and synthetic networks.
    
\end{abstract}

\title{Higher-order contagion processes in 3.99 dimensions}
\maketitle
Understanding the microscopic factors governing contagion processes is essential to monitor and forecast complex social dynamics \cite{Pastor2015}. For example, real contagion mechanisms are known to be shaped by an intricate mix of cultural \cite{centola2007} and behavioral aspects \cite{Funk2009}, along with the intrinsic origin of the initial ``spreaders'' \cite{Kitsak2010}.

In this context, recent studies have suggested that the inclusion of higher-order interactions --i.e., interactions involving more than two elements at a time-- in contagion dynamics models may also significantly influence phase transitions,  potentially triggering, e.g., discontinuous transitions \cite{reviewexplosive,Cai2015}. Thus, higher-order interactions have been evoked as the origin of a whole new panorama in dynamical processes on networks \cite{iacopini2019,Burgio2024,Ferraz2020,Kim2024,Battiston2021}. Hence, a crucial question arises: how do these new phenomena emerging from higher-order interactions relate to the classical theory of phase transitions?

In Kadanoff's words \cite{kadanoff1971}, two key ingredients define the critical behavior of a system \cite{MaBook,Binney,Amit,Zinn-Justin}: \emph{"All phase transition problems can be divided into a small number of different classes depending upon the dimensionality of the system and the symmetries of the ordered state"}. In the network context, dimensionality translates into the graph spectral dimension \cite{PRL2025,Burioni1996}. However, in contagion dynamics, the symmetries of the order state concern the evolution of the process, i.e., the time reversal invariance that characterizes possible future states \cite{Hinrichsen, MarroBook}, or so-called rapidity reversal in Reggeon field theory \cite{Ohtsuki1987, Henkel, Grassberger1979}.

\new{In this Letter, we build a bridge between the classical theory of phase transitions and higher-order mechanisms~\cite{Rosas2022} by mapping a recent simplicial contagion model~\cite{iacopini2019} onto a field-theoretical pairwise framework. Accordingly, the scope of our contribution is twofold: while the mesoscopic mapping is specific to simplicial contagion, the interplay between stochastic fluctuations and spectral dimension provides a broader criterion for assessing the stability of mean-field discontinuous transitions against fluctuation-induced rounding on networks. The mapping translates higher-order infection events into a mesoscopic Langevin description, where simplicial terms are absorbed into effective field coefficients and become formally equivalent to pairwise facilitation or threshold mechanisms. Within this setting, we show that 2-simplicial terms, i.e., triangles, are the relevant microscopic higher-order interactions able to drive first-order transitions, whereas higher-order terms are irrelevant at the mesoscopic level. We test the field description on different real contact networks and show that network spectral dimension controls how stochastic fluctuations constrain the critical properties of discontinuous transitions~\cite{imry1975}.}

\paragraph{\textbf{Field theoretical description.}}
Many agent-based contagion models are based on susceptible (S), infected (I), and recovered or susceptible (R/S) compartments, leading to the well-known SIS and SIR dynamics and their variants \cite{Radicchi2020,Bottcher}. In particular, systems with no immunization (i.e., with absorbing states), generally feature absorbing-active second-order phase transitions, belonging to the super-universal directed percolation universality class and being described by the paradigmatic contact process (CP) \cite{MAM1999,MAM1997,Henkel,BP}. Instead, those with immunization (no reinfection) belong to the so-called dynamic percolation universality class \cite{Grassberger1982,diSanto2017}.

\begin{figure}[hbtp]
    \centering
    \includegraphics[width=1.0\columnwidth]{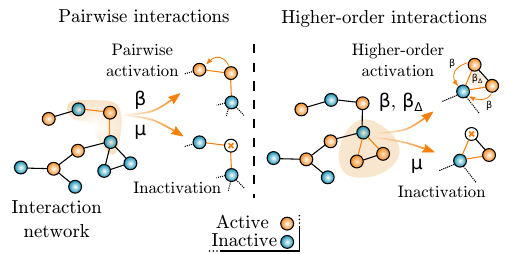}
    \caption{\textbf{Spreading dynamics.} Different microscopic rules are shown to account for pairwise interactions and higher-order (2-simplex) interactions. Different rates control the spreading process, namely: the activation rate ($\beta$), the inactivation rate ($\mu$), and the 2-simplex activation rate ($\beta_\Delta$).}
    \label{Sketch}
\end{figure}

Mathematically, SIS-like processes (see Fig.\ref{Sketch}) can be described using simple mean-field deterministic equations, such as the logistic one, which fully encodes the CP dynamics \cite{Harris} and reads $\dot{\rho}=-a\rho+b\rho^2$, where $\rho$ represents the fraction of infected sites in the system, $a$ is the growth rate ($a=\mu-\beta$), with $\beta$ ($\mu$) the corresponding microscopic activation/infection (inactivation/recovery) rates, as illustrated in Fig.\ref{Sketch}, and $b<0$ fixes the maximal activity density (e.g., carrying capacity) \cite{Murray}.


Additionally, the simplest non-equilibrium pairwise model with absorbing states showing a first-order/discontinuous transition is the quadratic contact process (QCP), where multiple contacts are required to get infected \cite{Bottcher, Ohtsuki1987, Villa2014,Elgart2006}. Note that in the QCP simultaneity is not necessary, provided there is enough long-term memory in multiple interactions \cite{Bizhani2012,Grassberger2016}.
This model includes facilitation terms with positive feedback mechanisms \cite{Eluding}, where the growth rate $a$ increases in the presence of activity, represented as $a\rightarrow a+\alpha \rho$, with $\alpha>0$. This change introduces a quadratic term of $-\alpha\rho^2/2$, which is analogous to substituting $b \rightarrow b-\alpha$ into the logistic equation while keeping the growth factor intact. This is equivalent to allowing activation or contagion in Fig.\ref{Sketch} if and only if at least $m\ge 2$ neighbors are infected. These specific terms can be derived from general microscopic interactions involving $l-$particle creation and $k-$particle annihilation \cite{Villa2014, Elgart2006}, leading to the following equation,
\begin{equation}
    \dot{\rho}=-a\rho+b\rho^2-c\rho^3,
\label{Langevin}
\end{equation}
where $b$ encodes facilitation effects and $c$ is a new cubic term that must be added to ensure a finite carrying capacity, preventing $\rho$ from diverging when $b>0$.

Aiming at fully capturing the relevant phenomenology, this equation can generally be extended to take into account the explicit network structure and/or demographic stochasticity. \new{Namely, the new equation reads,}
\begin{equation}
    \dot{\rho}_i=-a\rho_i+b\rho_i^2-c\rho_i^3-\sum_jL_{ij} \rho_j+\sqrt{\rho_i}\eta_i\left(t\right)\,
\label{Langevin-Sp}
\end{equation}
where $\eta(t)$ is a (zero mean, unit variance) Gaussian noise, and the spatial coupling appears through the Laplacian operator $L$, which governs diffusion, in continuum spaces simply becoming the differential operator $-\nabla^2$ or its regular lattice discrete counterpart.


Depending on the parameter values this very general equation --widely used to model ecological environments \cite{Eluding}-- can manifest two alternative scenarios: either a smooth (transcritical) or an abrupt (discontinuous) transition between an inactive and an active state. 
We emphasize that both scenarios depend on the parameter $b$, which controls the nature of the phase transition. In short, any value of $b>0$ results in a first-order (discontinuous) phase transition, while any value of $b<0$ will lead to a second-order (continuous) phase transition. Within this minimal coarse-grained description, the sign of $b$ determines the local bifurcation structure. However, in finite dimensions, fluctuations and correlations beyond this approximation may induce deviations from this criterion. The physical \new{meaning} of $b$ can be understood by considering a generic facilitation term in the underlying dynamics \cite{Villa2014, Eluding}, which is present in multiple systems: synaptic facilitation in neuroscience \cite{LG}, the Allee effect in ecology \cite{Taylor2005}, systems biology \cite{Pal2013}, or climate and social sciences \cite{Scheffer, Sole}. We also stress that from a phenomenological perspective, the previous mechanism can also be encoded in a general non-linear threshold function that leads to analogous first-order phase transitions \cite{LG} (see Supplemental Material (SM) \cite{SM} for an extended discussion on the issue).

We now show that the simplicial contagion model recently proposed in \cite{iacopini2019} can be mapped into Eq. \eqref{Langevin-Sp}. We first examine the mean-field (MF) equation stemming from the microscopic rates governing the so-called higher-order CP (see Fig.\ref{Sketch}). This MF equation can be appropriately rearranged in the form of Eq.~\ref{Langevin} to include the expansion up to 1-simplices (edges) and 2-simplices (triangles) into the corresponding coefficients, namely,
\begin{equation}
\dot{\rho}=(-\mu+\beta\langle\kappa\rangle)\rho+(\beta_\Delta\langle\kappa_\Delta\rangle-\beta\langle\kappa\rangle)\rho^2-\beta_\Delta\langle\kappa_\Delta\rangle\rho^3,
\label{HOrder}
\end{equation}
where $\mu$ is the inactivation (recovery) rate, $\beta$ is the activation (infection) rate, $\beta_\Delta$ the 2-simplex activation rate, $\langle \kappa \rangle$ the mean network connectivity and $\langle \kappa_\Delta \rangle$ the average number of 2-simplices incident on a node.

The first implication of this equation is that the order of the phase transition depends only on the relationship $(\beta_\Delta\langle\kappa_\Delta\rangle-\beta\langle\kappa\rangle)$. Therefore, higher-order interactions may act similarly to a local facilitation mechanism. Thus, the \emph{ad hoc} condition $\beta_\Delta\langle\kappa_\Delta\rangle>\beta\langle\kappa\rangle$, can be fine-tuned to induce the \emph{first-order} transition. However, the most significant consequence of the expansion in Eq.\eqref{HOrder} (see SM \cite{SM} for further terms) is that $\omega-$simplices contribute to the CP dynamics as $\mathcal{O}(\rho^\omega) \propto \rho^\omega (t) - \rho^{\omega+1}(t)$. Hence, simplices with $\omega>2$ are expected to be irrelevant \cite{Hinrichsen} (in the jargon of the Renormalization Group) at the level of the MF description. This suggests that heterogeneity in most real-world networks affects the location of the critical infection rate, while having little impact on the universal properties of the transition. In dense networks, which effectively correspond to infinite-dimensional systems, a similar critical behavior can be obtained after an appropriate rescaling of temporal scales.

We emphasize that Eq.\eqref{HOrder} is only strictly valid beyond the upper critical dimension. To consider finite spectral dimensions, the complete field-theoretical description at coarse-grained scales must consider the diffusive coupling term using the discrete Laplacian operator plus loop expansions, as established in standard Reggeon Field Theory \cite{Hinrichsen,Henkel}. The Laplacian term is the discrete counterpart of $\nabla^2\rho$, accounting for interactions between node neighbors, and describes dynamics on coarser scales, both in heterogeneous and homogeneous cases. Consequently, on these scales, the effects of the 2-simplex interaction, which affects the phase transition at a microscopic level, can be reabsorbed through a suitable value of the constant $b$, leading to a pairwise field-theoretical description. The new description now reads,
\begin{multline}
\dot{\rho_i}=(-\mu+\beta\langle\kappa\rangle)\rho_i+(\beta_\Delta\langle\kappa_\Delta\rangle-\beta\langle\kappa\rangle)\rho_i^2+\\-\beta_\Delta\langle\kappa_\Delta\rangle\rho_i^3-\sum_jL_{ij} \rho_j+\sqrt{\rho_i}\eta_i(t) .
\label{HOrder-Lapl}
\end{multline}
In what follows, we analyze how Eq.\eqref{HOrder-Lapl} can describe the nature of the phase transition when simulating the model described above.
\paragraph{\textbf{Detecting phase transitions on networks.}}
We now focus on detailed examples to test whether the mesoscopic coefficients related to the order parameter predict the nature of the phase transition. To this end, we examine the dependence of $b$ on different synthetic networks, where we control both the process generating them and any potential finite-size effect. We simulate the higher-order CP, with microscopic transition rates set to $\beta_\Delta=2\beta$ to enhance higher-order (discontinuous) effects during scaling analyses (see discussion in SM \cite{SM}). Specifically, we define $\lambda=\langle \kappa \rangle \frac{\beta}{\mu}$, fixing $\mu=0.05$. For consistency, we set $b$ as $b=2\langle\kappa_\Delta\rangle-\langle\kappa\rangle$ from now on.

\begin{figure}[hbtp]
    \centering
    \includegraphics[width=1.0\columnwidth]{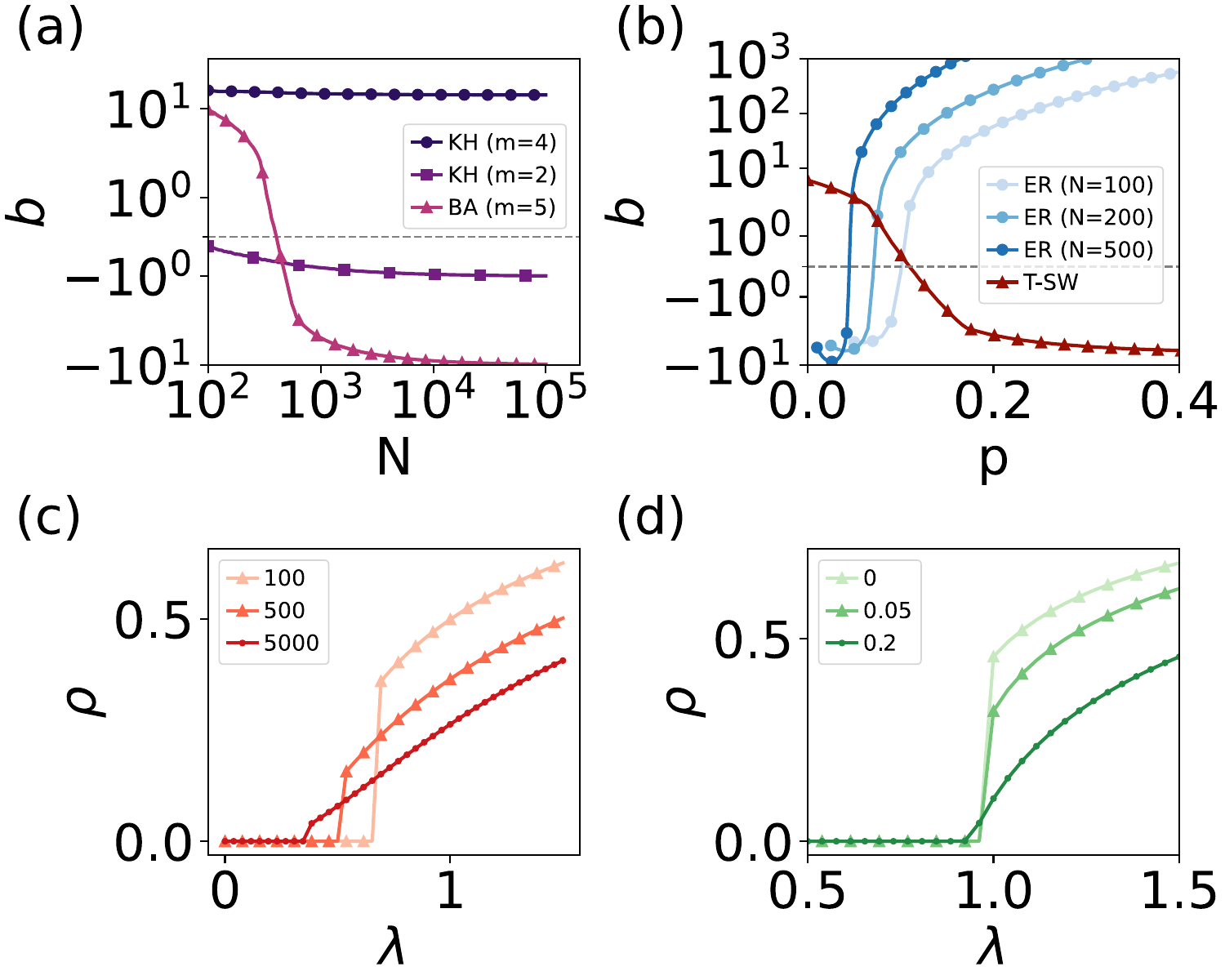}
    \caption{\textbf{Synthetic networks.} (a) Averaged value of b versus system size for different networks, BA with $m=5$, and KH with $m=4, ~p=1$, and $m=2,~p=0.5$ (see legend). (b) The averaged value of b versus rewiring probability for a 2D T-SW network (red triangles; $N=2450$) and the giant component of an ER network of different system sizes (see legend). Fraction of infected sites versus rescaled infection probability for (c) a BA network with $m=5$ and different system sizes (see legend) and (d) 2D T-SW networks with $N=2450$ and different rewiring probabilities, $p$ (see legend).}
    \label{beff}
\end{figure}

Figures \ref{beff}(a) illustrate how $b$ varies with the system size for different networks. Finite-size Barab\'asi-Albert (BA) networks show a size-dependent \new{continuous/discontinuous} phase transition for any $m\geq2$, where $m$ represents the number of edges of each new node when it is added to the network (see Fig.\ref{beff}(a) and SM \cite{SM}). This is due to the vanishing clustering coefficient that characterizes these specific architectures \cite{BARev}. In contrast, recently \new{characterized} scale-invariant networks \cite{PRL2025}, such as the Kim and Holme (KH) networks \cite{KH} exhibit a constant $b-$value, therefore leading \new{(at finite-size)} to a first or a second-order phase transition depending on the sign of b (see SM \cite{SM}). We have also analyzed the case of networks with varying rewiring or linking probabilities, such as small-world (T-SW) ones, where a fraction $p$ of links in a triangular lattice is rewired, or Erd\" {o}s-R\'enyi (ER) networks (see Fig.\ref{beff}(b)).

To validate our theoretical predictions, we perform extensive simulations. Fig. \ref{beff}(c) reveals how finite-size effects can influence the nature of the phase transition in a BA network, depending on system size. In contrast, Fig.\ref{beff}(d) shows how the discontinuous-to-continuous crossover in the phase transition is qualitatively consistent with the field-theoretical framework, when considering a T-SW network with rewiring.

\begin{figure}[hbtp]
    \centering
    \includegraphics[width=1.0\columnwidth]{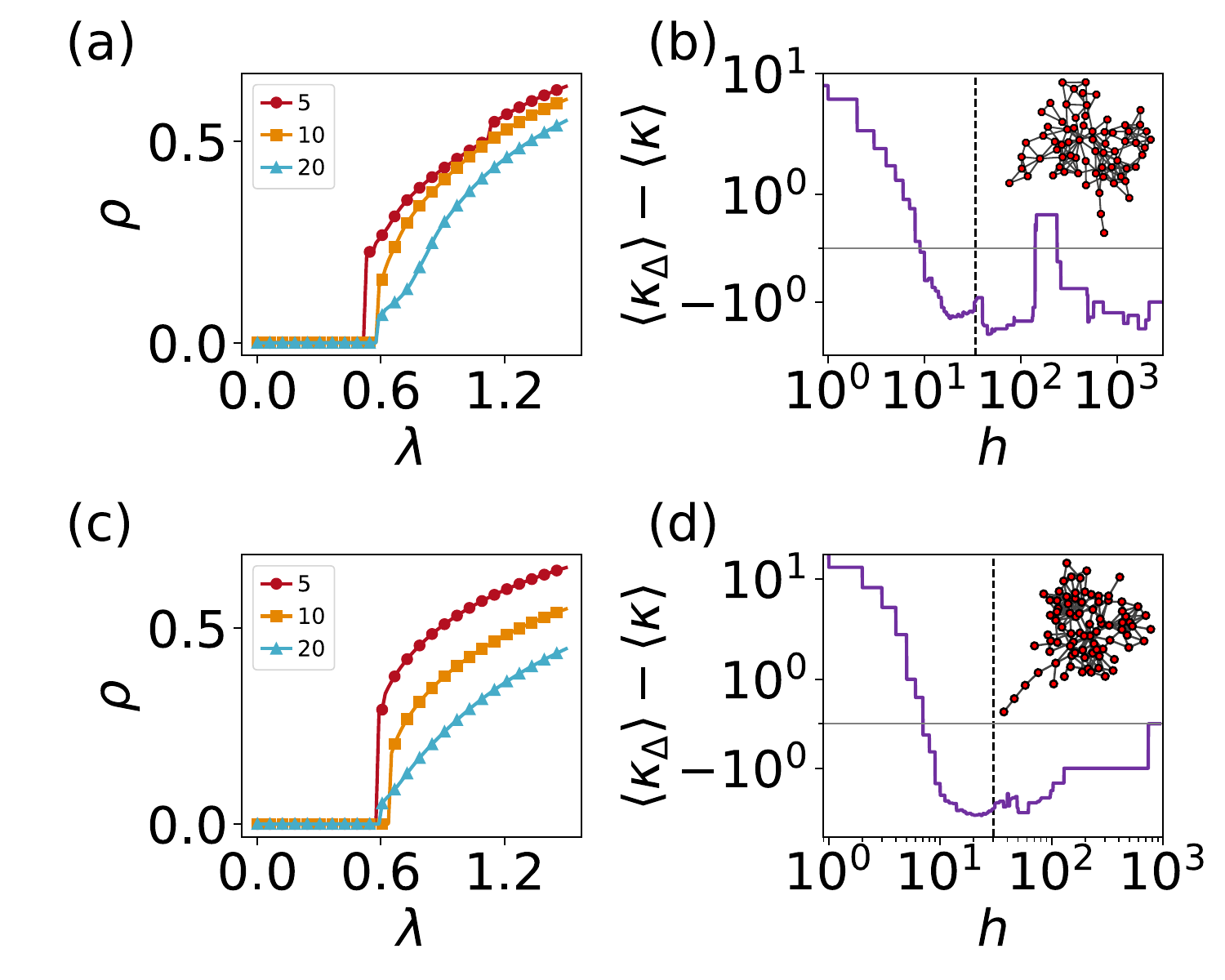}
    \caption{\textbf{Contact networks.}  Fraction of infected sites versus rescaled infection probability for contact networks in (a) a village in rural Malawi for different thresholds (see legend) and (c) in a workplace, both shown for varying threshold values (see legend). The difference between the average number of triangles $\langle \kappa_\Delta \rangle$ and the mean connectivity $\kappa$ as a function of the threshold is shown for (b) Malawi and (d) the workplace. Insets illustrate the resulting network for threshold value h = 10. The black dashed line marks where the giant component contains 50\% of the original nodes.}
    \label{RNets}
\end{figure}

We also validate our predictions by considering empirical social architectures. To this end, we consider publicly available data describing contact patterns based on proximity sensor technology collected by the SocioPatterns collaboration \cite{Sociopatterns}. These datasets contain the time-resolved interactions between individuals. The total interactions are aggregated, so that edges (connections) represent the duration of interactions. It is thus essential to select the relevant temporal scales used for aggregation to accurately assess their impact on contagion dynamics. To do that, we apply a threshold $h$ to filter out edges with low weights (see Fig.\ref{RNets}(a)), thereby extracting the backbone of such networks. As a technical remark, we ensure that, after thresholding, the networks still contain a giant component. \new{Therefore, the observed changes cannot be attributed to the tree-like critical structures that appear at the onset of percolation (see SM~\cite{SM}).}

In particular, we analyze data from a village in rural Malawi \cite{Ozella} and a workplace from the French Institute for Public Health Surveillance (InVS) \cite{Genois2} as case studies. We refer to SM \cite{SM} for a comprehensive analysis of other networks, which includes data from high \cite{Genois1, Fournet, Mastrandrea} and primary schools \cite{Genois1, Stehle}, a hospital, scientific conferences, and other workplaces \cite{Genois1}. Figures~\ref{RNets}(a) and (c) show how the phase transition changes as the threshold value $h$ is varied for the rural village and workplace networks. In both cases, the effective coefficient $b$ changes sign at approximately $h\simeq 7$ [see Figs.~\ref{RNets}(b) and (d)], while the thresholded network still retains a giant component. \new{Notably, this occurs even for $\beta_\Delta=2\beta$, a choice that enhances simplicial effects.} The observed behavior is qualitatively consistent with the mesoscopic prediction, although finite-size and fluctuation effects may lead to deviations for small network sizes.

\paragraph{\textbf{Interplay between noise and dimensionality in discontinuous absorbing-active transitions.}}

After analyzing mesoscopic coefficients, we focus on the role of noise and heterogeneity in phase transitions. We aim to demonstrate that the network \new{spectral} dimension is key for understanding how heterogeneity and fluctuations interact to produce nontrivial effects.  To do this, we use selected scale-invariant networks \cite{PRL2025} with a well-defined and finite spectral dimension, $d_S$ \cite{LRG,PRL2025}. Note that, for networks, $d_S$ effectively plays the role of the Euclidean dimension \cite{Cassi1992,PRL2025}, e.g., defining the Gaussian model in non-integer dimensions \cite{Burioni1996}. 

\begin{figure}[hbtp]
    \centering
    \includegraphics[width=1.0\columnwidth]{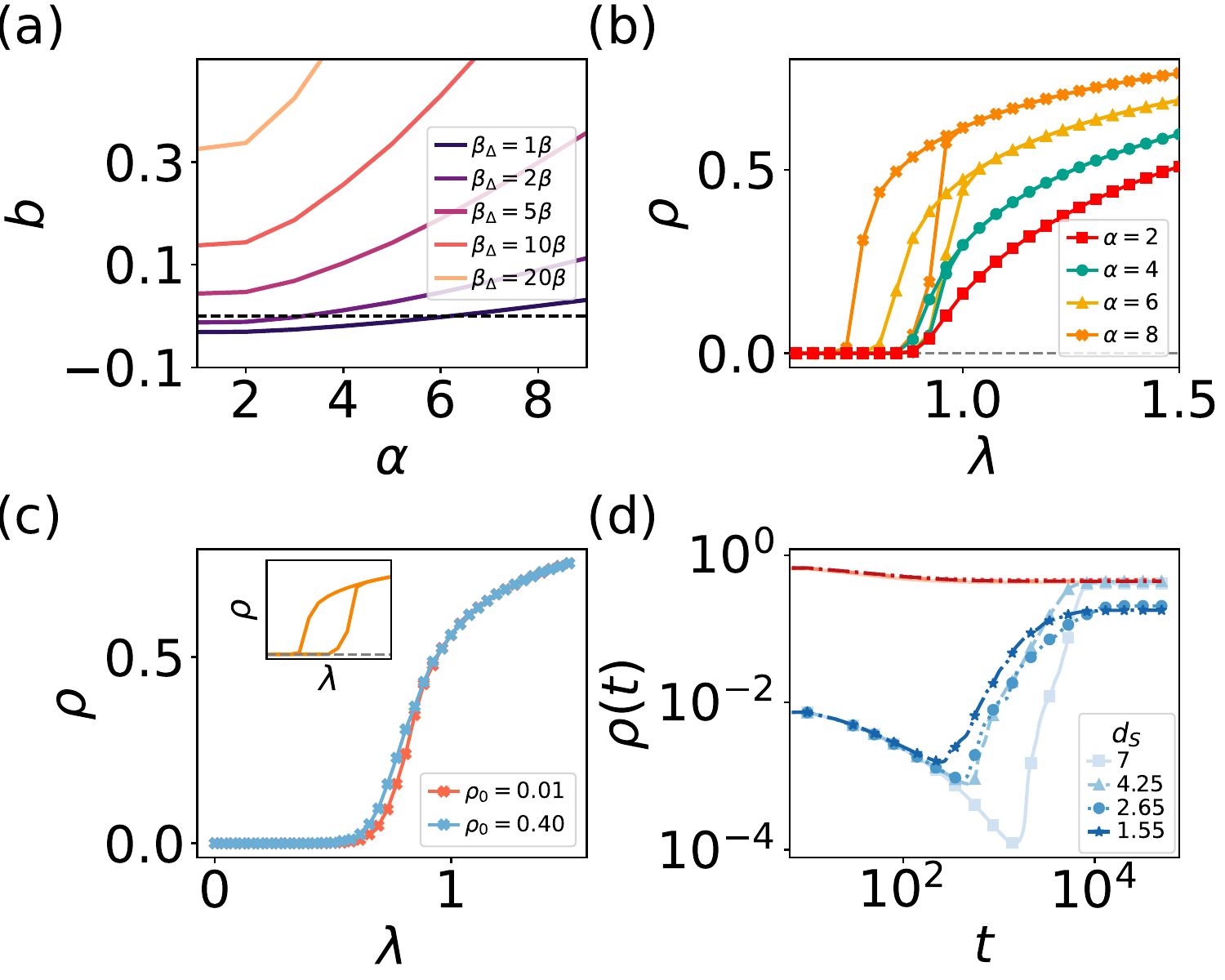}
    \caption{\textbf{Dimensionality and discontinuous transitions.} (a) Average effective value of $b$ versus $\alpha$ for HMNs with size $N=m_02^s$ for different rates $\beta_\Delta/\beta$ (see legend). Fraction of infected sites versus rescaled infection probability for (b) HMN with $m_0=3,$ and $s=11$ and different values of $\alpha$, for $\beta_\Delta=2\beta$. (c) HMN with $m_0=3,~s=15$  and $\alpha=8$ under the application of a small Gaussian variability $\mathcal{N}(\beta,\delta)$, with $\delta=0.05$. The inset shows the phase transition with $\delta=0$. (d) Temporal evolution of the density of infected nodes at criticality ($\lambda=0.36$) for CLRs obtaining different specified spectral dimensions (see legend). Beyond $d_s=4$, all networks display the same asymptotic critical scaling behavior. The corresponding effective values of $b$ are all positive ($b \gtrsim 3$), indicating that the system lies in the discontinuous regime. Red curves represent an initial density of infected nodes $\rho_0 = 0.5$ while blue ones $\rho_0= 5 \cdot 10^{-3}$. Parameters: $N=10^4$, $\tilde\mu\in[\mu-\delta,\mu+\delta]$, with $\delta=0.045$.}
    \label{HMNets}
\end{figure}

We first use hierarchical modular networks (HMNs) \cite{MorettiGP}, with a spectral dimension $d_S\in(1.25,2)$ \cite{PRL2025}, depending on specific parameters: $m_0$ (the nodes on basal modules) and $\alpha$ (controlling the density of links across scales, see SM \cite{SM}). As reported in Fig.\ref{HMNets}(a) and (b), the nature of the phase transition can be predicted by analyzing the sign of $b$, as in previous cases. Now, we introduce a \new{weak quenched disorder through small Gaussian fluctuations in the local infection rates}, $\beta$ and $\beta_\Delta$. Note that even small disorder amplitudes suppress the discontinuity, \new{preventing abrupt system-wide activation} across the entire population. \new{This behavior is consistent with previous results for regular lattices based on the Imry--Ma argument~\cite{imry1975,Villa2014}, and suggests that HMNs provide a heterogeneous-network realization of the same fluctuation-induced rounding mechanism.}

\new{To test this dimensional scenario in a controlled setting,} we analyze a modified version of 1D Chains with Long-Range interactions (CLR) \cite{Kotliar83}. This model has a continuously tunable spectral dimension, $d_s\in(1,\infty)$~\cite{Burioni97,Millan21} (see SM~\cite{SM}). Links between nodes $i$ and $j$ are drawn with probability decaying algebraically with their geometric distance, $p_{ij}\propto r_{ij}^{-(1+\sigma)}$, where $\sigma$ controls the effective dimensionality. We then introduce weak stochastic variability of amplitude $\delta$ in the recovery rate $\mu$. \new{This construction allows us to probe} how the discontinuous regime changes as the spectral dimension approaches the upper critical value $d_c=4$. For the CLR realizations shown in Fig.~\ref{HMNets}(d), the effective coefficient remains positive ($b\gtrsim 3$), so the mesoscopic theory places the system in the discontinuous regime. However, as $d_s$ approaches and exceeds $4$, hysteresis becomes progressively negligible, and trajectories initialized at low density relax toward the active branch, consistently with a mean-field-like stabilization of the discontinuous transition. \new{Thus, the CLR results support the interpretation that spectral dimension controls the relevance of fluctuations in absorbing-state transitions  (see also SM \cite{SM}).}

\paragraph{\textbf{Outlook.}} We have theoretically characterized the impact of \new{simplicial} higher-order interactions on phase transitions in contagion dynamics, in terms of the two classical ingredients characterizing critical phenomena: symmetries and dimensionality. Our results demonstrate that such interactions are formally equivalent to facilitation mechanisms encoded in the QCP. Therefore, the main consequence is that a field theory based on pairwise interactions remains valid to describe \new{this class of} contagion phenomena and governs dynamics at coarser scales. 

A by-product of our analysis is that only 2-simplex interactions are relevant and provide an additional microscopic mechanism able to generate first-order transitions. \new{We also provide evidence that spectral dimension controls the relevance of stochastic fluctuations in phase transitions on complex networks, with $d_{\rm crit}=4$ marking the expected crossover toward mean-field behavior.}

We stress that several metrics derived from the concept of clustering coefficient may show some correlation with the nature of the phase transition, as they depend on how the ratio between $\langle\kappa_\Delta\rangle$ and $\langle \kappa \rangle$ changes with system size. A recent example of this is the hyperedge overlap \cite{Malizia2025,Malizia2}, applied to 2-simplex interactions. However, these metrics cannot predict finite-size or dimensionality effects related to the overall network structure, as discussed in detail in the SM \cite{SM}. 

\new{The present theory provides a mesoscopic description of the simplicial higher-order contagion model~\cite{iacopini2019}. Our aim is not to predict model-specific details, such as microscopic thresholds, nor to establish an exact equivalence with arbitrary higher-order contagion rules. More general higher-order structures may involve additional descriptors beyond the average pairwise and simplicial connectivities considered here. Clarifying how these ingredients enter the effective theory is an important avenue for future analyses. We hope this work opens the door to future universal descriptions that link higher-order dynamics to classical universality classes in complex networks.}

\paragraph{\textbf{Acknowledgments.}}
S.M. personally thanks C. Strappaveccia for insightful and motivating discussions. We acknowledge P. Grassberger and M.A. Mu\~noz for very useful suggestions and comments. S.M. acknowledges support from the project `CODE - Coupling Opinion Dynamics with Epidemics', funded under PNRR Mission 4 `Education and Research' - Component C2 - Investment 1.1 - Next Generation EU `Fund for National Research Program and Projects of Significant National Interest' PRIN 2022 PNRR, grant code P2022AKRZ9. Partial support has also been received from the Agencia Estatal de Investigaci\'on and Fondo Europeo de Desarrollo Regional (FEDER, UE) under projects APASOS (PID2021-122256NB-C22), COSASTI (PID2024-157493NB-C22), and the Mar\'ia de Maeztu program, project CEX2021-001164-M, all funded by the  MCIN/AEI/10.13039/501100011033.
P.V. acknowledges the Spanish Ministry of Research and Innovation and Agencia Estatal de Investigaci\'on (AEI), MICIN/AEI/10.13039/501100011033, for financial support through Project PID2023-149174NB-I00, funded also by European Regional Development Funds, and  Ref. PID2020-113681GB-I00. 

\def\url#1{}

\begin{thebibliography}{62}%
\makeatletter
\providecommand \@ifxundefined [1]{%
 \@ifx{#1\undefined}
}%
\providecommand \@ifnum [1]{%
 \ifnum #1\expandafter \@firstoftwo
 \else \expandafter \@secondoftwo
 \fi
}%
\providecommand \@ifx [1]{%
 \ifx #1\expandafter \@firstoftwo
 \else \expandafter \@secondoftwo
 \fi
}%
\providecommand \natexlab [1]{#1}%
\providecommand \enquote  [1]{``#1''}%
\providecommand \bibnamefont  [1]{#1}%
\providecommand \bibfnamefont [1]{#1}%
\providecommand \citenamefont [1]{#1}%
\providecommand \href@noop [0]{\@secondoftwo}%
\providecommand \href [0]{\begingroup \@sanitize@url \@href}%
\providecommand \@href[1]{\@@startlink{#1}\@@href}%
\providecommand \@@href[1]{\endgroup#1\@@endlink}%
\providecommand \@sanitize@url [0]{\catcode `\\12\catcode `\$12\catcode
  `\&12\catcode `\#12\catcode `\^12\catcode `\_12\catcode `\%12\relax}%
\providecommand \@@startlink[1]{}%
\providecommand \@@endlink[0]{}%
\providecommand \url  [0]{\begingroup\@sanitize@url \@url }%
\providecommand \@url [1]{\endgroup\@href {#1}{\urlprefix }}%
\providecommand \urlprefix  [0]{URL }%
\providecommand \Eprint [0]{\href }%
\providecommand \doibase [0]{https://doi.org/}%
\providecommand \selectlanguage [0]{\@gobble}%
\providecommand \bibinfo  [0]{\@secondoftwo}%
\providecommand \bibfield  [0]{\@secondoftwo}%
\providecommand \translation [1]{[#1]}%
\providecommand \BibitemOpen [0]{}%
\providecommand \bibitemStop [0]{}%
\providecommand \bibitemNoStop [0]{.\EOS\space}%
\providecommand \EOS [0]{\spacefactor3000\relax}%
\providecommand \BibitemShut  [1]{\csname bibitem#1\endcsname}%
\let\auto@bib@innerbib\@empty
\bibitem [{\citenamefont {Pastor-Satorras}\ \emph {et~al.}(2015)\citenamefont
  {Pastor-Satorras}, \citenamefont {Castellano}, \citenamefont {Van~Mieghem},\
  and\ \citenamefont {Vespignani}}]{Pastor2015}%
  \BibitemOpen
  \bibfield  {author} {\bibinfo {author} {\bibfnamefont {R.}~\bibnamefont
  {Pastor-Satorras}}, \bibinfo {author} {\bibfnamefont {C.}~\bibnamefont
  {Castellano}}, \bibinfo {author} {\bibfnamefont {P.}~\bibnamefont
  {Van~Mieghem}},\ and\ \bibinfo {author} {\bibfnamefont {A.}~\bibnamefont
  {Vespignani}},\ }\href {https://doi.org/10.1103/RevModPhys.87.925} {\bibfield
   {journal} {\bibinfo  {journal} {Rev. Mod. Phys.}\ }\textbf {\bibinfo
  {volume} {87}},\ \bibinfo {pages} {925} (\bibinfo {year} {2015})}\BibitemShut
  {NoStop}%
\bibitem [{\citenamefont {Centola}\ \emph {et~al.}(2007)\citenamefont
  {Centola}, \citenamefont {Gonzalez-Avella}, \citenamefont {Eguiluz},\ and\
  \citenamefont {San~Miguel}}]{centola2007}%
  \BibitemOpen
  \bibfield  {author} {\bibinfo {author} {\bibfnamefont {D.}~\bibnamefont
  {Centola}}, \bibinfo {author} {\bibfnamefont {J.~C.}\ \bibnamefont
  {Gonzalez-Avella}}, \bibinfo {author} {\bibfnamefont {V.~M.}\ \bibnamefont
  {Eguiluz}},\ and\ \bibinfo {author} {\bibfnamefont {M.}~\bibnamefont
  {San~Miguel}},\ }\href {https://doi.org/10.1177/0022002707307632} {\bibfield
  {journal} {\bibinfo  {journal} {J. Confl. Resolut.}\ }\textbf {\bibinfo
  {volume} {51}},\ \bibinfo {pages} {905} (\bibinfo {year} {2007})}\BibitemShut
  {NoStop}%
\bibitem [{\citenamefont {Funk}\ \emph {et~al.}(2009)\citenamefont {Funk},
  \citenamefont {Gilad}, \citenamefont {Watkins},\ and\ \citenamefont
  {Jansen}}]{Funk2009}%
  \BibitemOpen
  \bibfield  {author} {\bibinfo {author} {\bibfnamefont {S.}~\bibnamefont
  {Funk}}, \bibinfo {author} {\bibfnamefont {E.}~\bibnamefont {Gilad}},
  \bibinfo {author} {\bibfnamefont {C.}~\bibnamefont {Watkins}},\ and\ \bibinfo
  {author} {\bibfnamefont {V.~A.}\ \bibnamefont {Jansen}},\ }\href
  {https://doi.org/10.1073/pnas.0810762106} {\bibfield  {journal} {\bibinfo
  {journal} {Proc. Natl. Acad. Sci. U.S.A.}\ }\textbf {\bibinfo {volume}
  {106}},\ \bibinfo {pages} {6872} (\bibinfo {year} {2009})}\BibitemShut
  {NoStop}%
\bibitem [{\citenamefont {Kitsak}\ and\ \citenamefont
  {et~al.}(2010)}]{Kitsak2010}%
  \BibitemOpen
  \bibfield  {author} {\bibinfo {author} {\bibfnamefont {M.}~\bibnamefont
  {Kitsak}}\ and\ \bibinfo {author} {\bibnamefont {et~al.}},\ }\href
  {https://doi.org/10.1038/nphys1746} {\bibfield  {journal} {\bibinfo
  {journal} {Nat. Phys.}\ }\textbf {\bibinfo {volume} {6}},\ \bibinfo {pages}
  {888} (\bibinfo {year} {2010})}\BibitemShut {NoStop}%
\bibitem [{\citenamefont {Raissa M.~D'Souza}\ and\ \citenamefont
  {Arenas}(2019)}]{reviewexplosive}%
  \BibitemOpen
  \bibfield  {author} {\bibinfo {author} {\bibfnamefont {J.~N.}\ \bibnamefont
  {Raissa M.~D'Souza}, \bibfnamefont {Jesus G\'omez-Garde\~nes}}\ and\ \bibinfo
  {author} {\bibfnamefont {A.}~\bibnamefont {Arenas}},\ }\href
  {https://doi.org/10.1080/00018732.2019.1650450} {\bibfield  {journal}
  {\bibinfo  {journal} {Adv. Phys.}\ }\textbf {\bibinfo {volume} {68}},\
  \bibinfo {pages} {123} (\bibinfo {year} {2019})}\BibitemShut {NoStop}%
\bibitem [{\citenamefont {Cai}\ \emph {et~al.}(2015)\citenamefont {Cai},
  \citenamefont {Chen}, \citenamefont {Ghanbarnejad},\ and\ \citenamefont
  {Grassberger}}]{Cai2015}%
  \BibitemOpen
  \bibfield  {author} {\bibinfo {author} {\bibfnamefont {W.}~\bibnamefont
  {Cai}}, \bibinfo {author} {\bibfnamefont {L.}~\bibnamefont {Chen}}, \bibinfo
  {author} {\bibfnamefont {F.}~\bibnamefont {Ghanbarnejad}},\ and\ \bibinfo
  {author} {\bibfnamefont {P.}~\bibnamefont {Grassberger}},\ }\href
  {https://doi.org/10.1038/nphys3457} {\bibfield  {journal} {\bibinfo
  {journal} {Nat. Phys.}\ }\textbf {\bibinfo {volume} {11}},\ \bibinfo {pages}
  {936} (\bibinfo {year} {2015})}\BibitemShut {NoStop}%
\bibitem [{\citenamefont {Iacopini}\ \emph {et~al.}(2019)\citenamefont
  {Iacopini}, \citenamefont {Petri}, \citenamefont {Barrat},\ and\
  \citenamefont {Latora}}]{iacopini2019}%
  \BibitemOpen
  \bibfield  {author} {\bibinfo {author} {\bibfnamefont {I.}~\bibnamefont
  {Iacopini}}, \bibinfo {author} {\bibfnamefont {G.}~\bibnamefont {Petri}},
  \bibinfo {author} {\bibfnamefont {A.}~\bibnamefont {Barrat}},\ and\ \bibinfo
  {author} {\bibfnamefont {V.}~\bibnamefont {Latora}},\ }\href
  {https://doi.org/10.1038/s41467-019-10431-6} {\bibfield  {journal} {\bibinfo
  {journal} {Nat. Commun.}\ }\textbf {\bibinfo {volume} {10}},\ \bibinfo
  {pages} {2485} (\bibinfo {year} {2019})}\BibitemShut {NoStop}%
\bibitem [{\citenamefont {Burgio}\ \emph {et~al.}(2024)\citenamefont {Burgio},
  \citenamefont {G\'omez},\ and\ \citenamefont {Arenas}}]{Burgio2024}%
  \BibitemOpen
  \bibfield  {author} {\bibinfo {author} {\bibfnamefont {G.}~\bibnamefont
  {Burgio}}, \bibinfo {author} {\bibfnamefont {S.}~\bibnamefont {G\'omez}},\
  and\ \bibinfo {author} {\bibfnamefont {A.}~\bibnamefont {Arenas}},\ }\href
  {https://doi.org/10.1103/PhysRevLett.132.077401} {\bibfield  {journal}
  {\bibinfo  {journal} {Phys. Rev. Lett.}\ }\textbf {\bibinfo {volume} {132}},\
  \bibinfo {pages} {077401} (\bibinfo {year} {2024})}\BibitemShut {NoStop}%
\bibitem [{\citenamefont {de~Arruda}\ \emph {et~al.}(2020)\citenamefont
  {de~Arruda}, \citenamefont {Petri},\ and\ \citenamefont
  {Moreno}}]{Ferraz2020}%
  \BibitemOpen
  \bibfield  {author} {\bibinfo {author} {\bibfnamefont {G.~F.}\ \bibnamefont
  {de~Arruda}}, \bibinfo {author} {\bibfnamefont {G.}~\bibnamefont {Petri}},\
  and\ \bibinfo {author} {\bibfnamefont {Y.}~\bibnamefont {Moreno}},\ }\href
  {https://doi.org/10.1103/PhysRevResearch.2.023032} {\bibfield  {journal}
  {\bibinfo  {journal} {Phys. Rev. Res.}\ }\textbf {\bibinfo {volume} {2}},\
  \bibinfo {pages} {023032} (\bibinfo {year} {2020})}\BibitemShut {NoStop}%
\bibitem [{\citenamefont {Kim}\ and\ \citenamefont {Goh}(2024)}]{Kim2024}%
  \BibitemOpen
  \bibfield  {author} {\bibinfo {author} {\bibfnamefont {J.-H.}\ \bibnamefont
  {Kim}}\ and\ \bibinfo {author} {\bibfnamefont {K.-I.}\ \bibnamefont {Goh}},\
  }\href {https://doi.org/10.1103/PhysRevLett.132.087401} {\bibfield  {journal}
  {\bibinfo  {journal} {Phys. Rev. Lett.}\ }\textbf {\bibinfo {volume} {132}},\
  \bibinfo {pages} {087401} (\bibinfo {year} {2024})}\BibitemShut {NoStop}%
\bibitem [{\citenamefont {Battiston}\ \emph {et~al.}(2021)\citenamefont
  {Battiston} \emph {et~al.}}]{Battiston2021}%
  \BibitemOpen
  \bibfield  {author} {\bibinfo {author} {\bibfnamefont {F.}~\bibnamefont
  {Battiston}} \emph {et~al.},\ }\href
  {https://doi.org/10.1038/s41567-021-01371-4} {\bibfield  {journal} {\bibinfo
  {journal} {Nat. Phys.}\ }\textbf {\bibinfo {volume} {17}},\ \bibinfo {pages}
  {1093} (\bibinfo {year} {2021})}\BibitemShut {NoStop}%
\bibitem [{\citenamefont {Kadanoff}(1971)}]{kadanoff1971}%
  \BibitemOpen
  \bibfield  {author} {\bibinfo {author} {\bibfnamefont {L.~P.}\ \bibnamefont
  {Kadanoff}},\ }in\ \href@noop {} {\emph {\bibinfo {booktitle} {Proceedings of
  the Enrico Fermi Summer School of Physics, Varenna 1970}}},\ \bibinfo
  {editor} {edited by\ \bibinfo {editor} {\bibfnamefont {M.~S.}\ \bibnamefont
  {Green}}}\ (\bibinfo  {publisher} {Academic Press},\ \bibinfo {address}
  {London and New York},\ \bibinfo {year} {1971})\BibitemShut {NoStop}%
\bibitem [{\citenamefont {Ma}(2018)}]{MaBook}%
  \BibitemOpen
  \bibfield  {author} {\bibinfo {author} {\bibfnamefont {S.-K.}\ \bibnamefont
  {Ma}},\ }\href@noop {} {\emph {\bibinfo {title} {Modern Theory of Critical
  Phenomena}}}\ (\bibinfo  {publisher} {Routledge},\ \bibinfo {address} {New
  York},\ \bibinfo {year} {2018})\BibitemShut {NoStop}%
\bibitem [{\citenamefont {Binney}(1992)}]{Binney}%
  \BibitemOpen
  \bibfield  {author} {\bibinfo {author} {\bibfnamefont {J.~J.}\ \bibnamefont
  {Binney}},\ }\href@noop {} {\emph {\bibinfo {title} {The Theory of Critical
  Phenomena: An Introduction to the Renormalization Group}}}\ (\bibinfo
  {publisher} {Oxford University Press},\ \bibinfo {address} {Oxford},\
  \bibinfo {year} {1992})\BibitemShut {NoStop}%
\bibitem [{\citenamefont {Amit}\ and\ \citenamefont
  {Martin-Mayor}(2005)}]{Amit}%
  \BibitemOpen
  \bibfield  {author} {\bibinfo {author} {\bibfnamefont {D.~J.}\ \bibnamefont
  {Amit}}\ and\ \bibinfo {author} {\bibfnamefont {V.}~\bibnamefont
  {Martin-Mayor}},\ }\href {https://doi.org/10.1142/5715} {\emph {\bibinfo
  {title} {Field Theory, the Renormalization Group, and Critical Phenomena}}},\
  \bibinfo {edition} {3rd}\ ed.\ (\bibinfo  {publisher} {World Scientific},\
  \bibinfo {address} {Singapore},\ \bibinfo {year} {2005})\BibitemShut
  {NoStop}%
\bibitem [{\citenamefont {Zinn-Justin}(2007)}]{Zinn-Justin}%
  \BibitemOpen
  \bibfield  {author} {\bibinfo {author} {\bibfnamefont {J.}~\bibnamefont
  {Zinn-Justin}},\ }\href
  {https://doi.org/10.1093/acprof:oso/9780199227198.001.0001} {\emph {\bibinfo
  {title} {Phase Transitions and Renormalization Group}}}\ (\bibinfo
  {publisher} {Oxford University Press},\ \bibinfo {year} {2007})\BibitemShut
  {NoStop}%
\bibitem [{\citenamefont {Poggialini}\ \emph {et~al.}(2025)\citenamefont
  {Poggialini}, \citenamefont {Villegas}, \citenamefont {Mu\~noz},\ and\
  \citenamefont {Gabrielli}}]{PRL2025}%
  \BibitemOpen
  \bibfield  {author} {\bibinfo {author} {\bibfnamefont {A.}~\bibnamefont
  {Poggialini}}, \bibinfo {author} {\bibfnamefont {P.}~\bibnamefont
  {Villegas}}, \bibinfo {author} {\bibfnamefont {M.~A.}\ \bibnamefont
  {Mu\~noz}},\ and\ \bibinfo {author} {\bibfnamefont {A.}~\bibnamefont
  {Gabrielli}},\ }\href {https://doi.org/10.1103/PhysRevLett.134.057401}
  {\bibfield  {journal} {\bibinfo  {journal} {Phys. Rev. Lett.}\ }\textbf
  {\bibinfo {volume} {134}},\ \bibinfo {pages} {057401} (\bibinfo {year}
  {2025})}\BibitemShut {NoStop}%
\bibitem [{\citenamefont {Burioni}\ and\ \citenamefont
  {Cassi}(1996)}]{Burioni1996}%
  \BibitemOpen
  \bibfield  {author} {\bibinfo {author} {\bibfnamefont {R.}~\bibnamefont
  {Burioni}}\ and\ \bibinfo {author} {\bibfnamefont {D.}~\bibnamefont
  {Cassi}},\ }\href {https://doi.org/10.1103/PhysRevLett.76.1091} {\bibfield
  {journal} {\bibinfo  {journal} {Phys. Rev. Lett.}\ }\textbf {\bibinfo
  {volume} {76}},\ \bibinfo {pages} {1091} (\bibinfo {year}
  {1996})}\BibitemShut {NoStop}%
\bibitem [{\citenamefont {Hinrichsen}(2000)}]{Hinrichsen}%
  \BibitemOpen
  \bibfield  {author} {\bibinfo {author} {\bibfnamefont {H.}~\bibnamefont
  {Hinrichsen}},\ }\href {https://doi.org/10.1080/00018730050198152} {\bibfield
   {journal} {\bibinfo  {journal} {Adv. Phys.}\ }\textbf {\bibinfo {volume}
  {49}},\ \bibinfo {pages} {815} (\bibinfo {year} {2000})}\BibitemShut
  {NoStop}%
\bibitem [{\citenamefont {Marro}\ and\ \citenamefont
  {Dickman}(1999)}]{MarroBook}%
  \BibitemOpen
  \bibfield  {author} {\bibinfo {author} {\bibfnamefont {J.}~\bibnamefont
  {Marro}}\ and\ \bibinfo {author} {\bibfnamefont {R.}~\bibnamefont
  {Dickman}},\ }\href@noop {} {\emph {\bibinfo {title} {Nonequilibrium Phase
  Transitions in Lattice Models}}},\ Collection Alea-Saclay: Monographs and
  Texts in Statistical Physics\ (\bibinfo  {publisher} {Cambridge University
  Press},\ \bibinfo {address} {Cambridge},\ \bibinfo {year} {1999})\BibitemShut
  {NoStop}%
\bibitem [{\citenamefont {Ohtsuki}\ and\ \citenamefont
  {Keyes}(1987)}]{Ohtsuki1987}%
  \BibitemOpen
  \bibfield  {author} {\bibinfo {author} {\bibfnamefont {T.}~\bibnamefont
  {Ohtsuki}}\ and\ \bibinfo {author} {\bibfnamefont {T.}~\bibnamefont
  {Keyes}},\ }\href {https://doi.org/10.1103/PhysRevA.35.2697} {\bibfield
  {journal} {\bibinfo  {journal} {Phys. Rev. A}\ }\textbf {\bibinfo {volume}
  {35}},\ \bibinfo {pages} {2697} (\bibinfo {year} {1987})}\BibitemShut
  {NoStop}%
\bibitem [{\citenamefont {Henkel}\ \emph {et~al.}(2008)\citenamefont {Henkel},
  \citenamefont {Hinrichsen},\ and\ \citenamefont {L{\"u}beck}}]{Henkel}%
  \BibitemOpen
  \bibfield  {author} {\bibinfo {author} {\bibfnamefont {M.}~\bibnamefont
  {Henkel}}, \bibinfo {author} {\bibfnamefont {H.}~\bibnamefont {Hinrichsen}},\
  and\ \bibinfo {author} {\bibfnamefont {S.}~\bibnamefont {L{\"u}beck}},\
  }\href {https://doi.org/10.1007/978-3-540-78886-0} {\emph {\bibinfo {title}
  {Non-Equilibrium Phase Transitions: Absorbing Phase Transitions}}},\
  Theoretical and Mathematical Physics\ (\bibinfo  {publisher} {Springer},\
  \bibinfo {address} {Berlin},\ \bibinfo {year} {2008})\BibitemShut {NoStop}%
\bibitem [{\citenamefont {Grassberger}\ and\ \citenamefont {de~la
  Torre}(1979)}]{Grassberger1979}%
  \BibitemOpen
  \bibfield  {author} {\bibinfo {author} {\bibfnamefont {P.}~\bibnamefont
  {Grassberger}}\ and\ \bibinfo {author} {\bibfnamefont {A.}~\bibnamefont
  {de~la Torre}},\ }\href {https://doi.org/10.1016/0003-4916(79)90207-0}
  {\bibfield  {journal} {\bibinfo  {journal} {Ann. Phys.}\ }\textbf {\bibinfo
  {volume} {122}},\ \bibinfo {pages} {373} (\bibinfo {year}
  {1979})}\BibitemShut {NoStop}%
\bibitem [{\citenamefont {Rosas}\ \emph {et~al.}(2022)\citenamefont {Rosas},
  \citenamefont {Mediano}, \citenamefont {Luppi} \emph {et~al.}}]{Rosas2022}%
  \BibitemOpen
  \bibfield  {author} {\bibinfo {author} {\bibfnamefont {F.~E.}\ \bibnamefont
  {Rosas}}, \bibinfo {author} {\bibfnamefont {P.~A.~M.}\ \bibnamefont
  {Mediano}}, \bibinfo {author} {\bibfnamefont {A.~I.}\ \bibnamefont {Luppi}},
  \emph {et~al.},\ }\href {https://doi.org/10.1038/s41567-022-01548-5}
  {\bibfield  {journal} {\bibinfo  {journal} {Nat. Phys.}\ }\textbf {\bibinfo
  {volume} {18}},\ \bibinfo {pages} {476} (\bibinfo {year} {2022})}\BibitemShut
  {NoStop}%
\bibitem [{\citenamefont {Imry}\ and\ \citenamefont {Ma}(1975)}]{imry1975}%
  \BibitemOpen
  \bibfield  {author} {\bibinfo {author} {\bibfnamefont {Y.}~\bibnamefont
  {Imry}}\ and\ \bibinfo {author} {\bibfnamefont {S.-k.}\ \bibnamefont {Ma}},\
  }\href {https://doi.org/10.1103/PhysRevLett.35.1399} {\bibfield  {journal}
  {\bibinfo  {journal} {Phys. Rev. Lett.}\ }\textbf {\bibinfo {volume} {35}},\
  \bibinfo {pages} {1399} (\bibinfo {year} {1975})}\BibitemShut {NoStop}%
\bibitem [{\citenamefont {Radicchi}\ \emph {et~al.}(2020)\citenamefont
  {Radicchi}, \citenamefont {Castellano}, \citenamefont {Flammini},
  \citenamefont {Mu{\~n}oz},\ and\ \citenamefont {Notarmuzi}}]{Radicchi2020}%
  \BibitemOpen
  \bibfield  {author} {\bibinfo {author} {\bibfnamefont {F.}~\bibnamefont
  {Radicchi}}, \bibinfo {author} {\bibfnamefont {C.}~\bibnamefont
  {Castellano}}, \bibinfo {author} {\bibfnamefont {A.}~\bibnamefont
  {Flammini}}, \bibinfo {author} {\bibfnamefont {M.~A.}\ \bibnamefont
  {Mu{\~n}oz}},\ and\ \bibinfo {author} {\bibfnamefont {D.}~\bibnamefont
  {Notarmuzi}},\ }\href {https://doi.org/10.1103/PhysRevResearch.2.033171}
  {\bibfield  {journal} {\bibinfo  {journal} {Phys. Rev. Res.}\ }\textbf
  {\bibinfo {volume} {2}},\ \bibinfo {pages} {033171} (\bibinfo {year}
  {2020})}\BibitemShut {NoStop}%
\bibitem [{\citenamefont {B{\"o}ttcher}\ \emph {et~al.}(2017)\citenamefont
  {B{\"o}ttcher}, \citenamefont {Nagler},\ and\ \citenamefont
  {Herrmann}}]{Bottcher}%
  \BibitemOpen
  \bibfield  {author} {\bibinfo {author} {\bibfnamefont {L.}~\bibnamefont
  {B{\"o}ttcher}}, \bibinfo {author} {\bibfnamefont {J.}~\bibnamefont
  {Nagler}},\ and\ \bibinfo {author} {\bibfnamefont {H.~J.}\ \bibnamefont
  {Herrmann}},\ }\href {https://doi.org/10.1103/PhysRevLett.118.088301}
  {\bibfield  {journal} {\bibinfo  {journal} {Phys. Rev. Lett.}\ }\textbf
  {\bibinfo {volume} {118}},\ \bibinfo {pages} {088301} (\bibinfo {year}
  {2017})}\BibitemShut {NoStop}%
\bibitem [{\citenamefont {Mu{\~n}oz}\ \emph {et~al.}(1999)\citenamefont
  {Mu{\~n}oz}, \citenamefont {Dickman}, \citenamefont {Vespignani},\ and\
  \citenamefont {Zapperi}}]{MAM1999}%
  \BibitemOpen
  \bibfield  {author} {\bibinfo {author} {\bibfnamefont {M.~A.}\ \bibnamefont
  {Mu{\~n}oz}}, \bibinfo {author} {\bibfnamefont {R.}~\bibnamefont {Dickman}},
  \bibinfo {author} {\bibfnamefont {A.}~\bibnamefont {Vespignani}},\ and\
  \bibinfo {author} {\bibfnamefont {S.}~\bibnamefont {Zapperi}},\ }\href
  {https://doi.org/10.1103/PhysRevE.59.6175} {\bibfield  {journal} {\bibinfo
  {journal} {Phys. Rev. E}\ }\textbf {\bibinfo {volume} {59}},\ \bibinfo
  {pages} {6175} (\bibinfo {year} {1999})}\BibitemShut {NoStop}%
\bibitem [{\citenamefont {Mu{\~n}oz}\ \emph {et~al.}(1997)\citenamefont
  {Mu{\~n}oz}, \citenamefont {Grinstein},\ and\ \citenamefont {Tu}}]{MAM1997}%
  \BibitemOpen
  \bibfield  {author} {\bibinfo {author} {\bibfnamefont {M.~A.}\ \bibnamefont
  {Mu{\~n}oz}}, \bibinfo {author} {\bibfnamefont {G.}~\bibnamefont
  {Grinstein}},\ and\ \bibinfo {author} {\bibfnamefont {Y.}~\bibnamefont
  {Tu}},\ }\href {https://doi.org/10.1103/PhysRevE.56.5101} {\bibfield
  {journal} {\bibinfo  {journal} {Phys. Rev. E}\ }\textbf {\bibinfo {volume}
  {56}},\ \bibinfo {pages} {5101} (\bibinfo {year} {1997})}\BibitemShut
  {NoStop}%
\bibitem [{\citenamefont {di~Santo}\ \emph
  {et~al.}(2017{\natexlab{a}})\citenamefont {di~Santo}, \citenamefont
  {Villegas}, \citenamefont {Burioni},\ and\ \citenamefont {Mu{\~n}oz}}]{BP}%
  \BibitemOpen
  \bibfield  {author} {\bibinfo {author} {\bibfnamefont {S.}~\bibnamefont
  {di~Santo}}, \bibinfo {author} {\bibfnamefont {P.}~\bibnamefont {Villegas}},
  \bibinfo {author} {\bibfnamefont {R.}~\bibnamefont {Burioni}},\ and\ \bibinfo
  {author} {\bibfnamefont {M.~A.}\ \bibnamefont {Mu{\~n}oz}},\ }\href
  {https://doi.org/10.1103/PhysRevE.95.032115} {\bibfield  {journal} {\bibinfo
  {journal} {Phys. Rev. E}\ }\textbf {\bibinfo {volume} {95}},\ \bibinfo
  {pages} {032115} (\bibinfo {year} {2017}{\natexlab{a}})}\BibitemShut
  {NoStop}%
\bibitem [{\citenamefont {Grassberger}(1981)}]{Grassberger1982}%
  \BibitemOpen
  \bibfield  {author} {\bibinfo {author} {\bibfnamefont {P.}~\bibnamefont
  {Grassberger}},\ }in\ \href {https://doi.org/10.1007/978-3-642-81778-6_49}
  {\emph {\bibinfo {booktitle} {Nonlinear Phenomena in Chemical Dynamics}}},\
  \bibinfo {editor} {edited by\ \bibinfo {editor} {\bibfnamefont
  {C.}~\bibnamefont {Vidal}}\ and\ \bibinfo {editor} {\bibfnamefont
  {A.}~\bibnamefont {Pacault}}}\ (\bibinfo  {publisher} {Springer Berlin
  Heidelberg},\ \bibinfo {address} {Berlin, Heidelberg},\ \bibinfo {year}
  {1981})\ pp.\ \bibinfo {pages} {262--262}\BibitemShut {NoStop}%
\bibitem [{\citenamefont {di~Santo}\ \emph
  {et~al.}(2017{\natexlab{b}})\citenamefont {di~Santo}, \citenamefont
  {Villegas}, \citenamefont {Burioni},\ and\ \citenamefont
  {Mu\~noz}}]{diSanto2017}%
  \BibitemOpen
  \bibfield  {author} {\bibinfo {author} {\bibfnamefont {S.}~\bibnamefont
  {di~Santo}}, \bibinfo {author} {\bibfnamefont {P.}~\bibnamefont {Villegas}},
  \bibinfo {author} {\bibfnamefont {R.}~\bibnamefont {Burioni}},\ and\ \bibinfo
  {author} {\bibfnamefont {M.~A.}\ \bibnamefont {Mu\~noz}},\ }\href
  {https://doi.org/10.1103/PhysRevE.95.032115} {\bibfield  {journal} {\bibinfo
  {journal} {Phys. Rev. E}\ }\textbf {\bibinfo {volume} {95}},\ \bibinfo
  {pages} {032115} (\bibinfo {year} {2017}{\natexlab{b}})}\BibitemShut
  {NoStop}%
\bibitem [{\citenamefont {Harris}(1974)}]{Harris}%
  \BibitemOpen
  \bibfield  {author} {\bibinfo {author} {\bibfnamefont {T.~E.}\ \bibnamefont
  {Harris}},\ }\href {https://doi.org/10.1214/aop/1176996493} {\bibfield
  {journal} {\bibinfo  {journal} {Ann. Probab.}\ }\textbf {\bibinfo {volume}
  {2}},\ \bibinfo {pages} {969} (\bibinfo {year} {1974})}\BibitemShut {NoStop}%
\bibitem [{\citenamefont {Murray}(2002)}]{Murray}%
  \BibitemOpen
  \bibfield  {author} {\bibinfo {author} {\bibfnamefont {J.~D.}\ \bibnamefont
  {Murray}},\ }\href {https://doi.org/10.1007/b98868} {\emph {\bibinfo {title}
  {Mathematical Biology I. An Introduction}}}\ (\bibinfo  {publisher}
  {Springer},\ \bibinfo {address} {New York},\ \bibinfo {year}
  {2002})\BibitemShut {NoStop}%
\bibitem [{\citenamefont {Villa~Mart{\'i}n}\ \emph {et~al.}(2014)\citenamefont
  {Villa~Mart{\'i}n}, \citenamefont {Bonachela},\ and\ \citenamefont
  {Mu{\~n}oz}}]{Villa2014}%
  \BibitemOpen
  \bibfield  {author} {\bibinfo {author} {\bibfnamefont {P.}~\bibnamefont
  {Villa~Mart{\'i}n}}, \bibinfo {author} {\bibfnamefont {J.~A.}\ \bibnamefont
  {Bonachela}},\ and\ \bibinfo {author} {\bibfnamefont {M.~A.}\ \bibnamefont
  {Mu{\~n}oz}},\ }\href {https://doi.org/10.1103/PhysRevE.89.012145} {\bibfield
   {journal} {\bibinfo  {journal} {Phys. Rev. E}\ }\textbf {\bibinfo {volume}
  {89}},\ \bibinfo {pages} {012145} (\bibinfo {year} {2014})}\BibitemShut
  {NoStop}%
\bibitem [{\citenamefont {Elgart}\ and\ \citenamefont
  {Kamenev}(2006)}]{Elgart2006}%
  \BibitemOpen
  \bibfield  {author} {\bibinfo {author} {\bibfnamefont {V.}~\bibnamefont
  {Elgart}}\ and\ \bibinfo {author} {\bibfnamefont {A.}~\bibnamefont
  {Kamenev}},\ }\href {https://doi.org/10.1103/PhysRevE.74.041101} {\bibfield
  {journal} {\bibinfo  {journal} {Phys. Rev. E}\ }\textbf {\bibinfo {volume}
  {74}},\ \bibinfo {pages} {041101} (\bibinfo {year} {2006})}\BibitemShut
  {NoStop}%
\bibitem [{\citenamefont {Bizhani}\ \emph {et~al.}(2012)\citenamefont
  {Bizhani}, \citenamefont {Paczuski},\ and\ \citenamefont
  {Grassberger}}]{Bizhani2012}%
  \BibitemOpen
  \bibfield  {author} {\bibinfo {author} {\bibfnamefont {G.}~\bibnamefont
  {Bizhani}}, \bibinfo {author} {\bibfnamefont {M.}~\bibnamefont {Paczuski}},\
  and\ \bibinfo {author} {\bibfnamefont {P.}~\bibnamefont {Grassberger}},\
  }\href {https://doi.org/10.1103/PhysRevE.86.011128} {\bibfield  {journal}
  {\bibinfo  {journal} {Phys. Rev. E}\ }\textbf {\bibinfo {volume} {86}},\
  \bibinfo {pages} {011128} (\bibinfo {year} {2012})}\BibitemShut {NoStop}%
\bibitem [{\citenamefont {Grassberger}\ \emph {et~al.}(2016)\citenamefont
  {Grassberger}, \citenamefont {Chen}, \citenamefont {Ghanbarnejad},\ and\
  \citenamefont {Cai}}]{Grassberger2016}%
  \BibitemOpen
  \bibfield  {author} {\bibinfo {author} {\bibfnamefont {P.}~\bibnamefont
  {Grassberger}}, \bibinfo {author} {\bibfnamefont {L.}~\bibnamefont {Chen}},
  \bibinfo {author} {\bibfnamefont {F.}~\bibnamefont {Ghanbarnejad}},\ and\
  \bibinfo {author} {\bibfnamefont {W.}~\bibnamefont {Cai}},\ }\href
  {https://doi.org/10.1103/PhysRevE.93.042316} {\bibfield  {journal} {\bibinfo
  {journal} {Phys. Rev. E}\ }\textbf {\bibinfo {volume} {93}},\ \bibinfo
  {pages} {042316} (\bibinfo {year} {2016})}\BibitemShut {NoStop}%
\bibitem [{\citenamefont {Mart\'in}\ \emph {et~al.}(2015)\citenamefont
  {Mart\'in}, \citenamefont {Bonachela}, \citenamefont {Levin},\ and\
  \citenamefont {Mu{\~n}oz}}]{Eluding}%
  \BibitemOpen
  \bibfield  {author} {\bibinfo {author} {\bibfnamefont {P.~V.}\ \bibnamefont
  {Mart\'in}}, \bibinfo {author} {\bibfnamefont {J.~A.}\ \bibnamefont
  {Bonachela}}, \bibinfo {author} {\bibfnamefont {S.~A.}\ \bibnamefont
  {Levin}},\ and\ \bibinfo {author} {\bibfnamefont {M.~A.}\ \bibnamefont
  {Mu{\~n}oz}},\ }\href {https://doi.org/10.1073/pnas.1414708112} {\bibfield
  {journal} {\bibinfo  {journal} {Proc. Natl. Acad. Sci. U.S.A.}\ }\textbf
  {\bibinfo {volume} {112}},\ \bibinfo {pages} {E1828} (\bibinfo {year}
  {2015})}\BibitemShut {NoStop}%
\bibitem [{\citenamefont {Di~Santo}\ \emph {et~al.}(2018)\citenamefont
  {Di~Santo}, \citenamefont {Villegas}, \citenamefont {Burioni},\ and\
  \citenamefont {Mu{\~n}oz}}]{LG}%
  \BibitemOpen
  \bibfield  {author} {\bibinfo {author} {\bibfnamefont {S.}~\bibnamefont
  {Di~Santo}}, \bibinfo {author} {\bibfnamefont {P.}~\bibnamefont {Villegas}},
  \bibinfo {author} {\bibfnamefont {R.}~\bibnamefont {Burioni}},\ and\ \bibinfo
  {author} {\bibfnamefont {M.~A.}\ \bibnamefont {Mu{\~n}oz}},\ }\href
  {https://doi.org/10.1073/pnas.1712989115} {\bibfield  {journal} {\bibinfo
  {journal} {Proc. Natl. Acad. Sci. U.S.A.}\ }\textbf {\bibinfo {volume}
  {115}},\ \bibinfo {pages} {E1356} (\bibinfo {year} {2018})}\BibitemShut
  {NoStop}%
\bibitem [{\citenamefont {Taylor}\ and\ \citenamefont
  {Hastings}(2005)}]{Taylor2005}%
  \BibitemOpen
  \bibfield  {author} {\bibinfo {author} {\bibfnamefont {C.~M.}\ \bibnamefont
  {Taylor}}\ and\ \bibinfo {author} {\bibfnamefont {A.}~\bibnamefont
  {Hastings}},\ }\href {https://doi.org/10.1111/j.1461-0248.2005.00787.x}
  {\bibfield  {journal} {\bibinfo  {journal} {Ecol. Lett.}\ }\textbf {\bibinfo
  {volume} {8}},\ \bibinfo {pages} {895} (\bibinfo {year} {2005})}\BibitemShut
  {NoStop}%
\bibitem [{\citenamefont {Pal}\ \emph {et~al.}(2013)\citenamefont {Pal},
  \citenamefont {Pal}, \citenamefont {Ghosh},\ and\ \citenamefont
  {Bose}}]{Pal2013}%
  \BibitemOpen
  \bibfield  {author} {\bibinfo {author} {\bibfnamefont {M.}~\bibnamefont
  {Pal}}, \bibinfo {author} {\bibfnamefont {A.~K.}\ \bibnamefont {Pal}},
  \bibinfo {author} {\bibfnamefont {S.}~\bibnamefont {Ghosh}},\ and\ \bibinfo
  {author} {\bibfnamefont {I.}~\bibnamefont {Bose}},\ }\href
  {https://doi.org/10.1088/1478-3975/10/3/036010} {\bibfield  {journal}
  {\bibinfo  {journal} {Phys. Biol.}\ }\textbf {\bibinfo {volume} {10}},\
  \bibinfo {pages} {036010} (\bibinfo {year} {2013})}\BibitemShut {NoStop}%
\bibitem [{\citenamefont {Scheffer}(2020)}]{Scheffer}%
  \BibitemOpen
  \bibfield  {author} {\bibinfo {author} {\bibfnamefont {M.}~\bibnamefont
  {Scheffer}},\ }\href@noop {} {\emph {\bibinfo {title} {Critical transitions
  in nature and society}}}\ (\bibinfo  {publisher} {Princeton University
  Press},\ \bibinfo {address} {Princeton},\ \bibinfo {year} {2020})\BibitemShut
  {NoStop}%
\bibitem [{\citenamefont {Sol\'e}\ and\ \citenamefont
  {Bascompte}(2006)}]{Sole}%
  \BibitemOpen
  \bibfield  {author} {\bibinfo {author} {\bibfnamefont {R.}~\bibnamefont
  {Sol\'e}}\ and\ \bibinfo {author} {\bibfnamefont {J.}~\bibnamefont
  {Bascompte}},\ }\href@noop {} {\emph {\bibinfo {title} {Self-Organization in
  Complex Ecosystems.}}}\ (\bibinfo  {publisher} {Princeton University Press},\
  \bibinfo {address} {Princeton},\ \bibinfo {year} {2006})\BibitemShut
  {NoStop}%
\bibitem [{SM()}]{SM}%
  \BibitemOpen
  \href@noop {} {}\bibinfo {note} {See Supplemental Material at [] for further
  details.}\BibitemShut {Stop}%
\bibitem [{\citenamefont {Albert}\ and\ \citenamefont
  {Barab\'asi}(2002)}]{BARev}%
  \BibitemOpen
  \bibfield  {author} {\bibinfo {author} {\bibfnamefont {R.}~\bibnamefont
  {Albert}}\ and\ \bibinfo {author} {\bibfnamefont {A.-L.}\ \bibnamefont
  {Barab\'asi}},\ }\href {https://doi.org/10.1103/RevModPhys.74.47} {\bibfield
  {journal} {\bibinfo  {journal} {Rev. Mod. Phys.}\ }\textbf {\bibinfo {volume}
  {74}},\ \bibinfo {pages} {47} (\bibinfo {year} {2002})}\BibitemShut {NoStop}%
\bibitem [{\citenamefont {Holme}\ and\ \citenamefont {Kim}(2002)}]{KH}%
  \BibitemOpen
  \bibfield  {author} {\bibinfo {author} {\bibfnamefont {P.}~\bibnamefont
  {Holme}}\ and\ \bibinfo {author} {\bibfnamefont {B.~J.}\ \bibnamefont
  {Kim}},\ }\href {https://doi.org/10.1103/PhysRevE.65.026107} {\bibfield
  {journal} {\bibinfo  {journal} {Phys. Rev. E}\ }\textbf {\bibinfo {volume}
  {65}},\ \bibinfo {pages} {026107} (\bibinfo {year} {2002})}\BibitemShut
  {NoStop}%
\bibitem [{Soc()}]{Sociopatterns}%
  \BibitemOpen
  \href@noop {} {}\bibinfo {note} {SocioPatterns Collaboration.
  http://www.sociopatterns.org/. Accessed Dec 2024}\BibitemShut {NoStop}%
\bibitem [{\citenamefont {Ozella}\ \emph {et~al.}(2021)\citenamefont {Ozella},
  \citenamefont {Paolotti}, \citenamefont {Lichand}, \citenamefont
  {Rodr{\'\i}guez}, \citenamefont {Haenni}, \citenamefont {Phuka},
  \citenamefont {Leal-Neto},\ and\ \citenamefont {Cattuto}}]{Ozella}%
  \BibitemOpen
  \bibfield  {author} {\bibinfo {author} {\bibfnamefont {L.}~\bibnamefont
  {Ozella}}, \bibinfo {author} {\bibfnamefont {D.}~\bibnamefont {Paolotti}},
  \bibinfo {author} {\bibfnamefont {G.}~\bibnamefont {Lichand}}, \bibinfo
  {author} {\bibfnamefont {J.~P.}\ \bibnamefont {Rodr{\'\i}guez}}, \bibinfo
  {author} {\bibfnamefont {S.}~\bibnamefont {Haenni}}, \bibinfo {author}
  {\bibfnamefont {J.}~\bibnamefont {Phuka}}, \bibinfo {author} {\bibfnamefont
  {O.~B.}\ \bibnamefont {Leal-Neto}},\ and\ \bibinfo {author} {\bibfnamefont
  {C.}~\bibnamefont {Cattuto}},\ }\href
  {https://doi.org/10.1140/epjds/s13688-021-00302-w} {\bibfield  {journal}
  {\bibinfo  {journal} {EPJ Data Sci.}\ }\textbf {\bibinfo {volume} {10}},\
  \bibinfo {pages} {46} (\bibinfo {year} {2021})}\BibitemShut {NoStop}%
\bibitem [{\citenamefont {G{\'e}nois}\ \emph {et~al.}(2015)\citenamefont
  {G{\'e}nois}, \citenamefont {Vestergaard}, \citenamefont {Fournet},
  \citenamefont {Panisson}, \citenamefont {Bonmarin},\ and\ \citenamefont
  {Barrat}}]{Genois2}%
  \BibitemOpen
  \bibfield  {author} {\bibinfo {author} {\bibfnamefont {M.}~\bibnamefont
  {G{\'e}nois}}, \bibinfo {author} {\bibfnamefont {C.~L.}\ \bibnamefont
  {Vestergaard}}, \bibinfo {author} {\bibfnamefont {J.}~\bibnamefont
  {Fournet}}, \bibinfo {author} {\bibfnamefont {A.}~\bibnamefont {Panisson}},
  \bibinfo {author} {\bibfnamefont {I.}~\bibnamefont {Bonmarin}},\ and\
  \bibinfo {author} {\bibfnamefont {A.}~\bibnamefont {Barrat}},\ }\href
  {https://doi.org/10.1017/nws.2015.10} {\bibfield  {journal} {\bibinfo
  {journal} {Netw. Sci.}\ }\textbf {\bibinfo {volume} {3}},\ \bibinfo {pages}
  {326} (\bibinfo {year} {2015})}\BibitemShut {NoStop}%
\bibitem [{\citenamefont {G{\'e}nois}\ and\ \citenamefont
  {Barrat}(2018)}]{Genois1}%
  \BibitemOpen
  \bibfield  {author} {\bibinfo {author} {\bibfnamefont {M.}~\bibnamefont
  {G{\'e}nois}}\ and\ \bibinfo {author} {\bibfnamefont {A.}~\bibnamefont
  {Barrat}},\ }\href {https://doi.org/10.1140/epjds/s13688-018-0140-1}
  {\bibfield  {journal} {\bibinfo  {journal} {EPJ Data Sci.}\ }\textbf
  {\bibinfo {volume} {7}},\ \bibinfo {pages} {1} (\bibinfo {year}
  {2018})}\BibitemShut {NoStop}%
\bibitem [{\citenamefont {Fournet}\ and\ \citenamefont
  {Barrat}(2014)}]{Fournet}%
  \BibitemOpen
  \bibfield  {author} {\bibinfo {author} {\bibfnamefont {J.}~\bibnamefont
  {Fournet}}\ and\ \bibinfo {author} {\bibfnamefont {A.}~\bibnamefont
  {Barrat}},\ }\href {https://doi.org/10.1371/journal.pone.0107878} {\bibfield
  {journal} {\bibinfo  {journal} {PloS ONE}\ }\textbf {\bibinfo {volume} {9}},\
  \bibinfo {pages} {e107878} (\bibinfo {year} {2014})}\BibitemShut {NoStop}%
\bibitem [{\citenamefont {Mastrandrea}\ \emph {et~al.}(2015)\citenamefont
  {Mastrandrea}, \citenamefont {Fournet},\ and\ \citenamefont
  {Barrat}}]{Mastrandrea}%
  \BibitemOpen
  \bibfield  {author} {\bibinfo {author} {\bibfnamefont {R.}~\bibnamefont
  {Mastrandrea}}, \bibinfo {author} {\bibfnamefont {J.}~\bibnamefont
  {Fournet}},\ and\ \bibinfo {author} {\bibfnamefont {A.}~\bibnamefont
  {Barrat}},\ }\href {https://doi.org/10.1371/journal.pone.0136497} {\bibfield
  {journal} {\bibinfo  {journal} {PloS ONE}\ }\textbf {\bibinfo {volume}
  {10}},\ \bibinfo {pages} {e0136497} (\bibinfo {year} {2015})}\BibitemShut
  {NoStop}%
\bibitem [{\citenamefont {Stehl{\'e}}\ \emph {et~al.}(2011)\citenamefont
  {Stehl{\'e}} \emph {et~al.}}]{Stehle}%
  \BibitemOpen
  \bibfield  {author} {\bibinfo {author} {\bibfnamefont {J.}~\bibnamefont
  {Stehl{\'e}}} \emph {et~al.},\ }\href
  {https://doi.org/10.1371/journal.pone.0023176} {\bibfield  {journal}
  {\bibinfo  {journal} {PloS ONE}\ }\textbf {\bibinfo {volume} {6}},\ \bibinfo
  {pages} {e23176} (\bibinfo {year} {2011})}\BibitemShut {NoStop}%
\bibitem [{\citenamefont {Cassi}(1992)}]{Cassi1992}%
  \BibitemOpen
  \bibfield  {author} {\bibinfo {author} {\bibfnamefont {D.}~\bibnamefont
  {Cassi}},\ }\href {https://doi.org/10.1103/PhysRevLett.68.3631} {\bibfield
  {journal} {\bibinfo  {journal} {Phys. Rev. Lett.}\ }\textbf {\bibinfo
  {volume} {68}},\ \bibinfo {pages} {3631} (\bibinfo {year}
  {1992})}\BibitemShut {NoStop}%
\bibitem [{\citenamefont {Villegas}\ \emph {et~al.}(2023)\citenamefont
  {Villegas}, \citenamefont {Gili}, \citenamefont {Caldarelli},\ and\
  \citenamefont {Gabrielli}}]{LRG}%
  \BibitemOpen
  \bibfield  {author} {\bibinfo {author} {\bibfnamefont {P.}~\bibnamefont
  {Villegas}}, \bibinfo {author} {\bibfnamefont {T.}~\bibnamefont {Gili}},
  \bibinfo {author} {\bibfnamefont {G.}~\bibnamefont {Caldarelli}},\ and\
  \bibinfo {author} {\bibfnamefont {A.}~\bibnamefont {Gabrielli}},\ }\href
  {https://doi.org/10.1038/s41567-022-01866-8} {\bibfield  {journal} {\bibinfo
  {journal} {Nat. Phys.}\ }\textbf {\bibinfo {volume} {19}},\ \bibinfo {pages}
  {445} (\bibinfo {year} {2023})}\BibitemShut {NoStop}%
\bibitem [{\citenamefont {Moretti}\ and\ \citenamefont
  {Mu{\~n}oz}(2013)}]{MorettiGP}%
  \BibitemOpen
  \bibfield  {author} {\bibinfo {author} {\bibfnamefont {P.}~\bibnamefont
  {Moretti}}\ and\ \bibinfo {author} {\bibfnamefont {M.~A.}\ \bibnamefont
  {Mu{\~n}oz}},\ }\href {https://doi.org/10.1038/ncomms3521} {\bibfield
  {journal} {\bibinfo  {journal} {Nat. Comm.}\ }\textbf {\bibinfo {volume}
  {4}},\ \bibinfo {pages} {2521} (\bibinfo {year} {2013})}\BibitemShut
  {NoStop}%
\bibitem [{\citenamefont {Kotliar}\ \emph {et~al.}(1983)\citenamefont
  {Kotliar}, \citenamefont {Anderson},\ and\ \citenamefont
  {Stein}}]{Kotliar83}%
  \BibitemOpen
  \bibfield  {author} {\bibinfo {author} {\bibfnamefont {G.}~\bibnamefont
  {Kotliar}}, \bibinfo {author} {\bibfnamefont {P.~W.}\ \bibnamefont
  {Anderson}},\ and\ \bibinfo {author} {\bibfnamefont {D.~L.}\ \bibnamefont
  {Stein}},\ }\href {https://doi.org/10.1103/PhysRevB.27.602} {\bibfield
  {journal} {\bibinfo  {journal} {Phys. Rev. B}\ }\textbf {\bibinfo {volume}
  {27}},\ \bibinfo {pages} {602} (\bibinfo {year} {1983})}\BibitemShut
  {NoStop}%
\bibitem [{\citenamefont {Burioni}\ and\ \citenamefont
  {Cassi}(1997)}]{Burioni97}%
  \BibitemOpen
  \bibfield  {author} {\bibinfo {author} {\bibfnamefont {R.}~\bibnamefont
  {Burioni}}\ and\ \bibinfo {author} {\bibfnamefont {D.}~\bibnamefont
  {Cassi}},\ }\href {https://doi.org/10.1142/S0217984997001316} {\bibfield
  {journal} {\bibinfo  {journal} {Mod. Phys. Lett. B}\ }\textbf {\bibinfo
  {volume} {11}},\ \bibinfo {pages} {1095} (\bibinfo {year}
  {1997})}\BibitemShut {NoStop}%
\bibitem [{\citenamefont {Mill\'an}\ \emph {et~al.}(2021)\citenamefont
  {Mill\'an}, \citenamefont {Gori}, \citenamefont {Battiston}, \citenamefont
  {Enss},\ and\ \citenamefont {Defenu}}]{Millan21}%
  \BibitemOpen
  \bibfield  {author} {\bibinfo {author} {\bibfnamefont {A.~P.}\ \bibnamefont
  {Mill\'an}}, \bibinfo {author} {\bibfnamefont {G.}~\bibnamefont {Gori}},
  \bibinfo {author} {\bibfnamefont {F.}~\bibnamefont {Battiston}}, \bibinfo
  {author} {\bibfnamefont {T.}~\bibnamefont {Enss}},\ and\ \bibinfo {author}
  {\bibfnamefont {N.}~\bibnamefont {Defenu}},\ }\href
  {https://doi.org/10.1103/PhysRevResearch.3.023015} {\bibfield  {journal}
  {\bibinfo  {journal} {Phys. Rev. Res.}\ }\textbf {\bibinfo {volume} {3}},\
  \bibinfo {pages} {023015} (\bibinfo {year} {2021})}\BibitemShut {NoStop}%
\bibitem [{\citenamefont {Malizia}\ \emph
  {et~al.}(2025{\natexlab{a}})\citenamefont {Malizia}, \citenamefont
  {Lamata-Ot\'{i}n}, \citenamefont {Frasca} \emph {et~al.}}]{Malizia2025}%
  \BibitemOpen
  \bibfield  {author} {\bibinfo {author} {\bibfnamefont {F.}~\bibnamefont
  {Malizia}}, \bibinfo {author} {\bibfnamefont {S.}~\bibnamefont
  {Lamata-Ot\'{i}n}}, \bibinfo {author} {\bibfnamefont {M.}~\bibnamefont
  {Frasca}}, \emph {et~al.},\ }\href
  {https://doi.org/10.1038/s41467-024-55506-1} {\bibfield  {journal} {\bibinfo
  {journal} {Nat. Commun.}\ }\textbf {\bibinfo {volume} {16}},\ \bibinfo
  {pages} {555} (\bibinfo {year} {2025}{\natexlab{a}})}\BibitemShut {NoStop}%
\bibitem [{\citenamefont {Malizia}\ \emph
  {et~al.}(2025{\natexlab{b}})\citenamefont {Malizia}, \citenamefont
  {Guzm\'{a}n}, \citenamefont {Iacopini},\ and\ \citenamefont
  {Kiss}}]{Malizia2}%
  \BibitemOpen
  \bibfield  {author} {\bibinfo {author} {\bibfnamefont {F.}~\bibnamefont
  {Malizia}}, \bibinfo {author} {\bibfnamefont {A.}~\bibnamefont {Guzm\'{a}n}},
  \bibinfo {author} {\bibfnamefont {I.}~\bibnamefont {Iacopini}},\ and\
  \bibinfo {author} {\bibfnamefont {I.~Z.}\ \bibnamefont {Kiss}},\ }\href
  {https://arxiv.org/abs/2501.17800} {\bibfield  {journal} {\bibinfo  {journal}
  {arXiv preprint}\ } (\bibinfo {year} {2025}{\natexlab{b}})},\ \Eprint
  {https://arxiv.org/abs/2501.17800} {arXiv:2501.17800 [physics.soc-ph]}
  \BibitemShut {NoStop}%
\end{thebibliography}
%

\clearpage


\section*{Supplementary material}

# Supplementary Information: Higher-order contagion processes in 1.99 dimensions

Sandro Meloni,[1, 2, 3, *] Andrea Gabrielli,[3, 4, 5] and Pablo Villegas[3, 6, †]

[1]*Institute for Cross-Disciplinary Physics and Complex Systems (IFISC), CSIC-UIB, 07122 - Palma de Mallorca, Spain*
[2]*Institute for Applied Mathematics Mauro Picone (IAC) CNR, Via dei Taurini 19, 00185 - Rome, Italy*
[3]*'Enrico Fermi' Research Center (CREF), Via Panisperna 89A, 00184 - Rome, Italy*
[4]*Dipartimento di Ingegneria Civile, Informatica e delle Tecnologie Aeronautiche,*
*Università degli Studi "Roma Tre", Via Vito Volterra 62, 00146 - Rome, Italy.*
[5]*Istituto dei Sistemi Complessi (ISC) - CNR, Rome, Italy.*
[6]*Instituto Carlos I de Física Teórica y Computacional, Univ. de Granada, E-18071, Granada, Spain.*

## CONTENTS

---

[*] sandromeloni@cnr.it
[†] pablo.villegas@cref.it

## I. FIELD DESCRIPTION OF THE SIMPLICIAL CONTAGION MODEL

Given the set of infection probabilities $\beta = \beta_\omega, \omega = 1, ..., D$ and a recovery probability $\mu$, it is possible to derive the mean-field expression for the temporal evolution of the density of infected nodes $\rho(t)$ for the simplicial contagion model as explained in [1], which leads to the expression,

$$\dot\rho = -u\rho(t) + \sum_{\omega=1}^{D} \beta_\omega \langle \kappa_\omega \rangle \rho^\omega(t)[1-\rho(t)] \qquad (1)$$

where, for each $\omega = 1, ..., D$, $\kappa_i(\omega)$ is the generalized (simplicial) degree of a 0-dimensional face (node i), i.e., the number of $\omega$-dimensional simplices incident to the node i, and $\langle \kappa_\omega \rangle$ is its average over all the nodes.

As we know, terms over $\rho^3$ are irrelevant for describing the contact process or SIS-like models (i.e., models belonging to the directed percolation universality class) in mean-field [2]. Thus, expanding the previous equation up to order $\rho^3$, we can derive the following description, including noise and spatial coupling effects,

$$\dot\rho_i = (-\mu + \beta\langle\kappa\rangle)\rho_i + (\beta_\Delta\langle\kappa_\Delta\rangle - \beta\langle\kappa\rangle)\rho_i^2 - (\beta_\Delta\langle\kappa_\Delta\rangle - \beta_\square\langle\kappa_\square\rangle)\rho_i^3 - \sum_j L_{ij}\rho_j + \sqrt{\rho_i}\eta_i(t), \qquad (2)$$

where now $\beta_\square$ represents the 3-simplex activation rate.

Note that $\beta_\square$ only affects the carrying capacity term, while any higher-order term $\omega > 3$ will strictly vanish and is irrelevant to the characterization of the problem. Consequently, we have fixed it to zero for simplicity in all the manuscript.

## II. THE RESCALING OF INFECTION RATES.

As originally described, the normalization parameters of the simplicial contagion model (SCM) [1] is defined as a function of the following rescaled microscopic rates,

$$\begin{cases} \beta = \frac{\mu}{\langle\kappa\rangle}\lambda \\ \beta_\Delta = \frac{\mu}{\langle\kappa_\Delta\rangle}\lambda_\Delta \end{cases} \qquad (3)$$

This effectively changes the corresponding values of $\beta$ and $\beta_\Delta$ (that is, the microscopic infection rates) in cases where the clustering coefficient goes to zero. Figure 1(a) shows the expected behavior of the ratio between $\beta$ and $\beta_\Delta$ for the case of a BA network with $m = 5$ in case the microscopic rates are rescaled as indicated in Eq.(3). This leads to the unphysical effect of modifying the ratio between microscopic interaction rates as the network grows, thus altering the $\rho^2$ term in Eq.(2). Instead, we select the value of $\beta$, making $\beta_\Delta$ scale as a constant factor $\alpha$, which ensures the microscopic rates remain stable in the infinite size limit (note that the opposite choice gives the same results if one selects to fix $\beta_\Delta$ and scale $\beta$ as a function of it). Figure 1(b) shows the rates $\beta$ and $\beta_\Delta$ as a function of the system size for BA networks with $m = 5$.

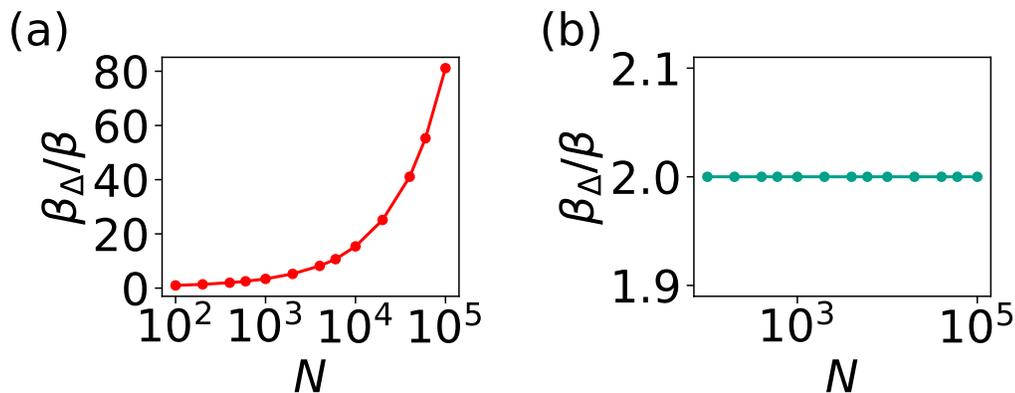

FIG. 1. Ratio between $\beta_\Delta$ and $\beta$ for a BA network with $m = 5$ by rescaling (a) both with $\langle\kappa\rangle$ and $\langle\kappa_\Delta\rangle$ and (b) only with $\langle\kappa\rangle$. Curves have been averaged over $10^2$ network realizations.

## III. THRESHOLD FUNCTIONS AS A MESOSCOPIC MECHANISM TO GENERATE FIRST-ORDER PHASE TRANSITIONS

Let us now discuss one of the simplest possible mesoscopic mechanisms leading to a first-order phase transition: an activity threshold encoded in a sigmoid (transduction) function [3]. This is described by the following Langevin equation, which defines the dynamics of an average firing rate or global activity $\rho$,

$$\dot{\rho} = -a\rho + (1-\rho)\tanh(R\rho - \Theta) + \nabla^2\rho + \sqrt{\rho}\eta(t). \qquad (4)$$

This characterizes, among others, the Wilson-Cowan model for neural network dynamics in its simplest form, interpreting $R$ as a constant representing synaptic strength and $\Theta$ as a threshold value that can be fixed to unity, being both regulated by a sigmoid (transduction) function.

Hence, by Taylor-expanding Eq.(4) close to the transition point, this leads us to,

$$\dot{\rho} = -\tanh(\Theta) + \left(-a + R\left(1 + \tanh^2(\Theta)\right) + \tanh(\Theta)\right)\rho + \left(R\left(\tanh^2(\Theta) - 1\right) + R^2\left(\tanh(\Theta) - \tanh^3(\Theta)\right)\right)\rho^2$$
$$- \left(\frac{1}{3}\left(3\tanh^4(\Theta) - 4\tanh^2(\Theta) + 1\right)R^3 + \left(-3\tanh^3(\Theta) + 3\tanh(\Theta)\right)R^2\right)\rho^3 + \mathcal{O}\left(\rho^4\right) + \nabla^2\rho + \sqrt{\rho}\eta(t) \qquad (5)$$

We can thus carefully inspect Eq.(5) to compare the corresponding coefficients described in the main text to characterize the order of the phase transition. In particular, we obtain,

$$\begin{cases} b = & \left(R\left(\tanh^2(\Theta) - 1\right) + R^2\left(\tanh(\Theta) - \tanh^3(\Theta)\right)\right) \\ c = & \left(\tanh^4(\Theta) - \frac{4}{3}\tanh^2(\Theta) + 1\right)R^3 \\ & + \left(-\tanh^3(\Theta) + \tanh(\Theta)\right)R^2 \end{cases} \qquad (6)$$

Figure 3 shows the different possible phase transitions when considering a threshold function on the general Langevin equation of Eq.(4) depending on the corresponding signs of $b$ and $c$.

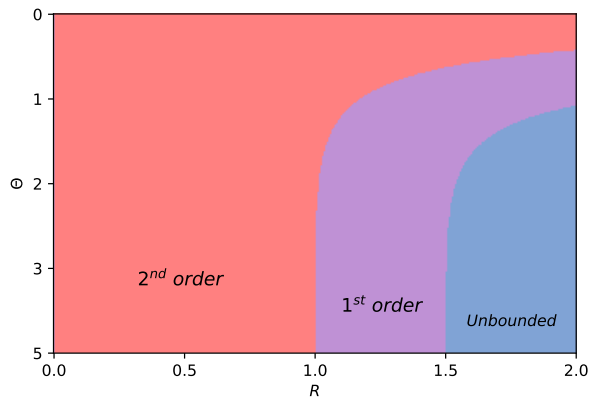

FIG. 2. $\theta$-$R$ phase diagram showing the order of the phase transition as expected from the analysis of $b$ and $c$ coefficients for the stochastic Langevin equation defined in Eq.(4). This gives place to different phase transitions of first and second order as a function of the sign of $b$. The unbounded region involves an unphysical value of $c$.

## IV. KIM AND HOLME NETWORKS

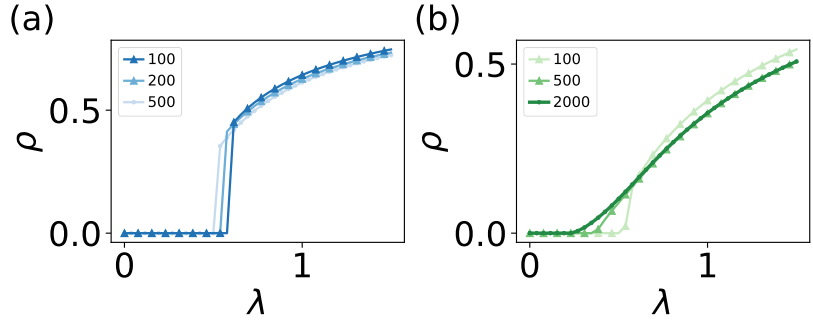

FIG. 3. Fraction of infected sites versus rescaled infection probability for the KH networks indicated in the main text with (a) $m = 4, p = 1$ (b>0), and (b) $m = 2, p = 0.5$ (b<0).

## V. HIERARCHIC MODULAR NETWORKS (HMN)

We have also analyzed the synthetic hierarchical networks originally described in [4]. These networks, called hierarchic modular networks (HMNs), have been specifically generated to closely resemble the structure of real brain networks. In particular, HMN consists of N nodes or neurons and L links or synapses, organized into hierarchical levels for easy analysis. The HMN model that we exemplify here uses a bottom-to-top approach. First, we construct local fully connected moduli and group them recursively by establishing new inter-moduli links in a deterministic manner with a level-dependent number of connections (HMN).

Hence, the growing algorithm works as follows, as detailed in [4]:

(i) At each hierarchical level $l = 1, 2, ..., s$, different pairs of blocks are selected, each with size $2^{i-1}m_0$. All possible undirected $4^{i-1}m_0^2$ connections between the two blocks are evaluated and established to avoid repetitions.

(ii) The number of connections between blocks at each level is set a priori at a constant value $\alpha$.

(iii) This method is stochastic in assigning connections, although the number of them (as well as the degree of the network) is fixed deterministically, being,

$$\langle \kappa \rangle = m_0 - 1 + \frac{2\alpha}{m_0}(1 - 2^{-s}).$$

Figure 4 shows the expected value of $b = 2\langle \kappa_\Delta \rangle - \langle \kappa \rangle$ for different system sizes, $N$. Note that, in principle, for lower values of $m_0$, the phase transition is of second order (negative values of b), while it should become of first order for higher values of $\alpha$. However, as explained in the main text, the low spectral dimension plays a crucial role in these specific networks, allowing them to avoid explosive phase transitions in the thermodynamic limit.

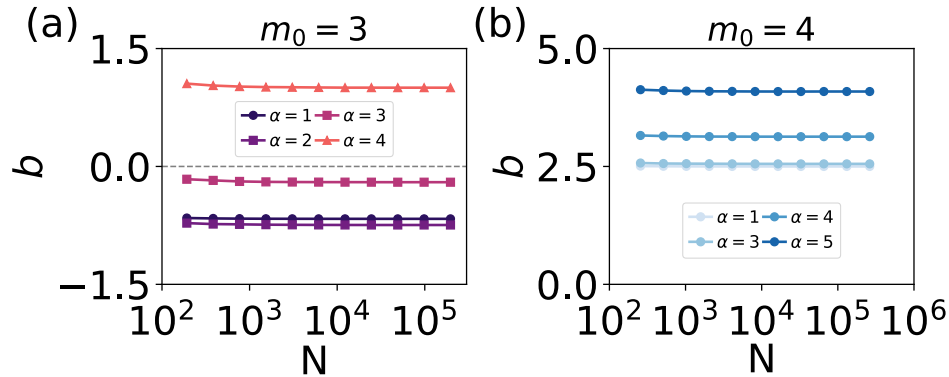

FIG. 4. Analysis of $b$, defined as $2\langle \kappa_\Delta \rangle - \langle \kappa \rangle$, versus system size, $N$, for different hierarchic modular networks with different values of $\alpha$ of (a) $m_0 = 3$ and (b) $m_0 = 4$. All curves have been averaged over $10^3$ independent realizations.

## A. Phase transitions in HMN networks

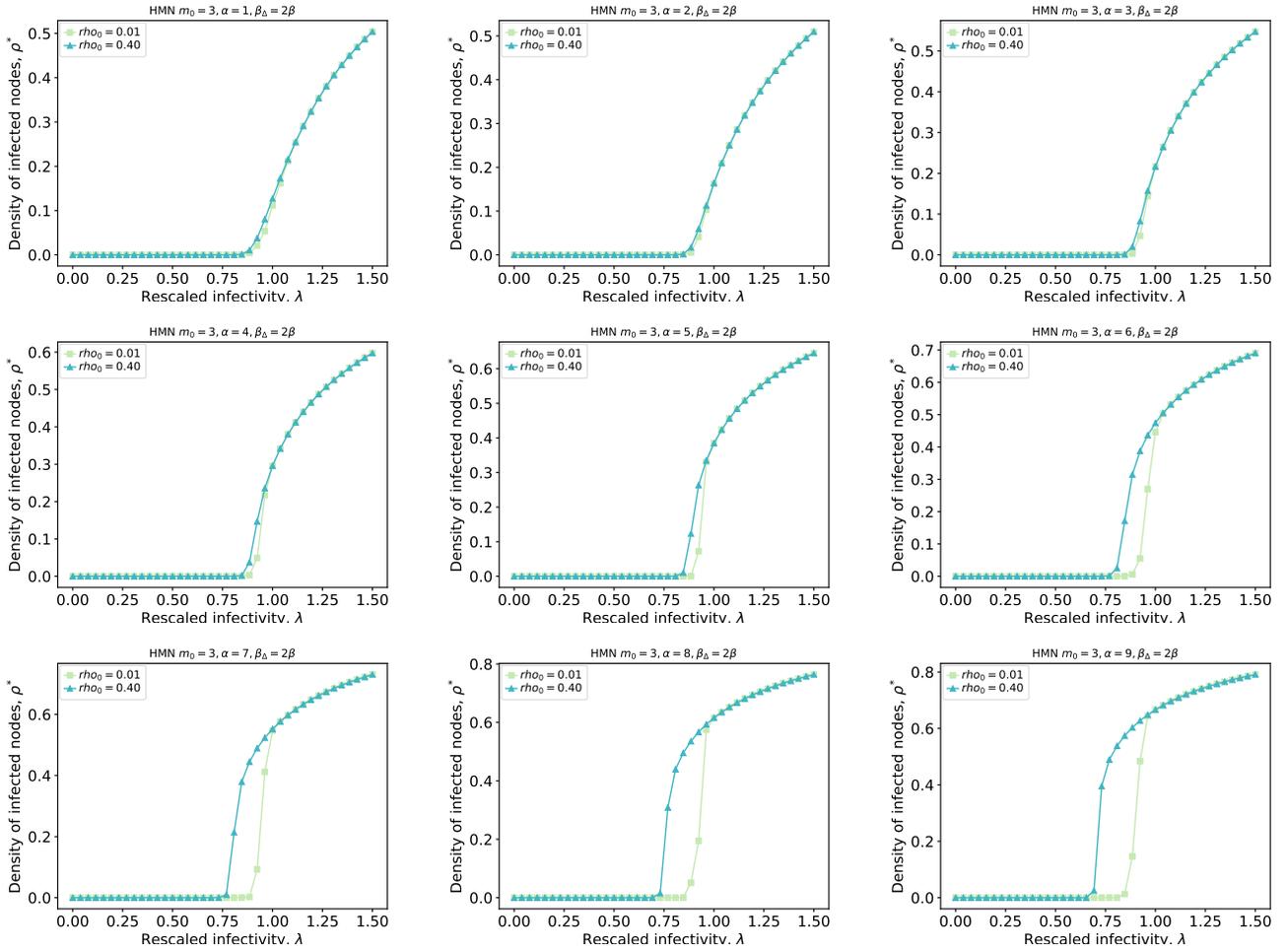

FIG. 5. Fraction of infected sites versus rescaled infection probability for HMN networks with $m_0 = 3$ varying the value of $\alpha$. Note the emergence of a first-order phase transition as predicted in the main text. Parameters: $s = 11$ hierarchical levels.

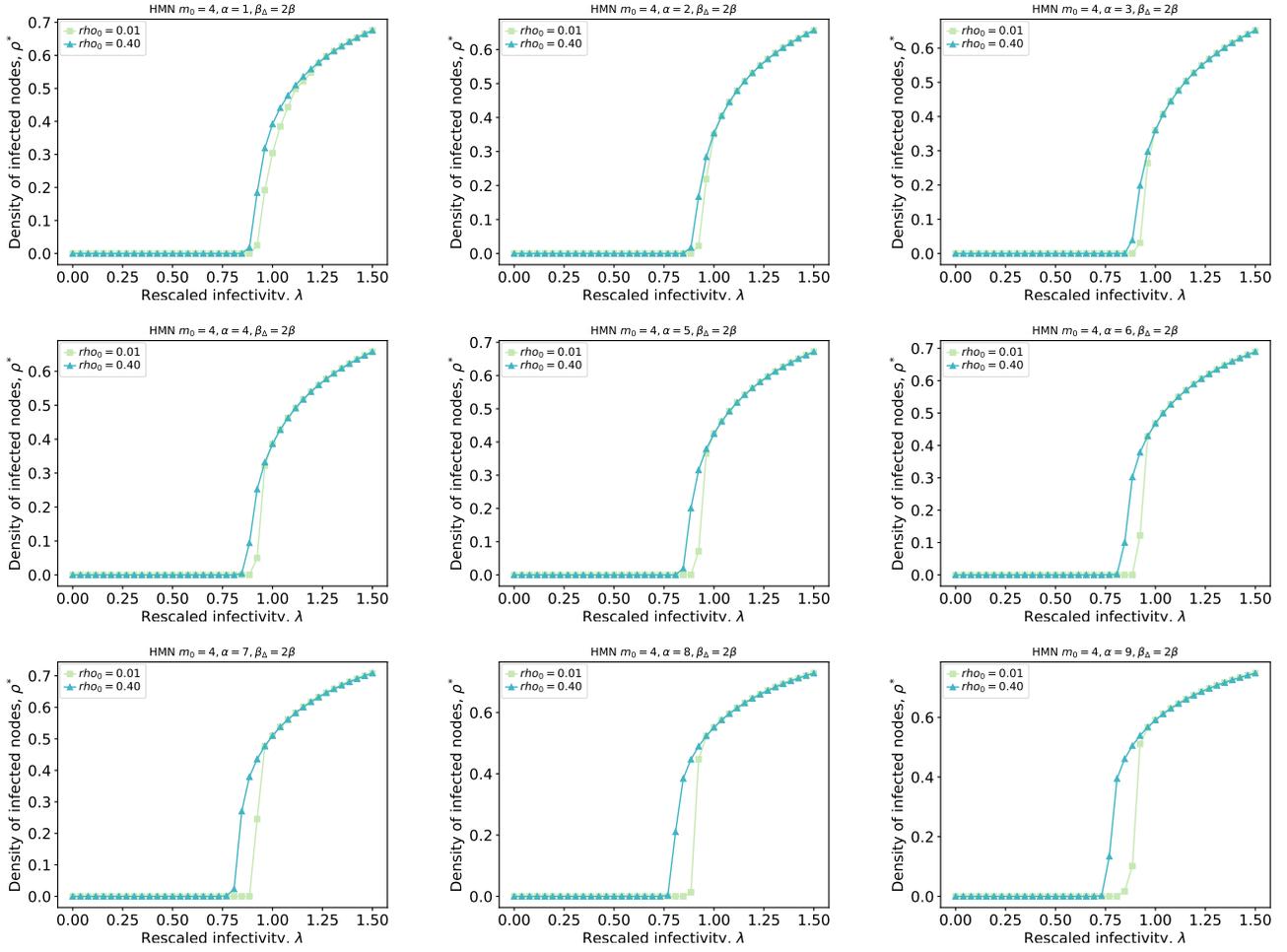

FIG. 6. Fraction of infected sites versus rescaled infection probability for HMN networks with $m_0 = 4$ varying the value of $\alpha$. Note the emergence of a first-order phase transition as predicted in the main text. Parameters: $s = 11$ hierarchical levels.

6

## VI. ANALYSIS OF REAL CONTAGION NETWORKS

For completeness, we present here a complete analysis of different real contagion networks from the SocioPatterns database at http://www.sociopatterns.org. All the datasets comprise weighted networks where contact event between individuals is defined as a set of successive time windows during which the individuals are detected in contact. In particular, we have considered the whole contact time as the weight of the network. Thus, by thresholding the network using a specific weight $h$, we can produce sparser versions of every network. Figure 7 shows the giant cluster of the network ($P_\infty$), and the relative number of edges ($E_\infty$) as a function of the threshold. We have also computed the value of $\langle \kappa_\Delta \rangle - \langle \kappa \rangle$ (violet lines in Fig. 7), which governs the phase transition of Eq. (2), as explained in the main text. In particular, Figure 7 comprises real data for various social contexts: a workplace, with data collected in two different years (InVS13, InVS15), a hospital (LH10), a primary school (LyonSchool), a scientific conference (SFHH) and a high school (Thiers13), from [5]. As can be appreciated for all these networks, after breaking them (vertical dashed lines), they are always strongly connected networks that will lead to first-order phase transitions.

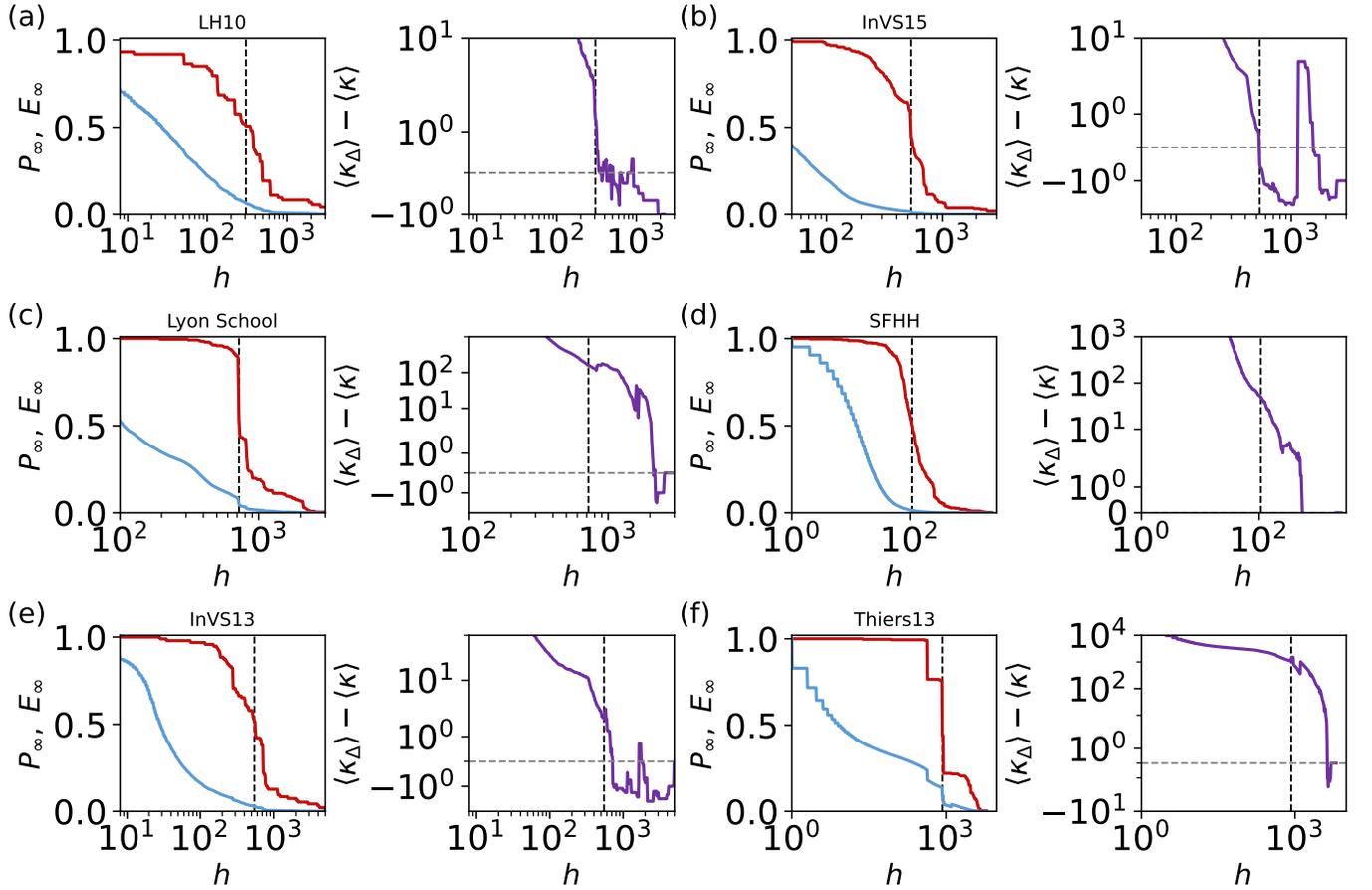

FIG. 7. **Real networks.** (Left) Giant cluster of the network ($P_\infty$) (red line) and relative number of edges, $E_\infty$, (blue line) versus threshold weight $h$. The black dashed line shows the value $h$ for which the giant component of the network has only 50% of the original number of individuals. (Right) $\langle \kappa_\Delta \rangle - \langle \kappa \rangle$ (violet line) versus threshold weight $h$. The black dashed line represents the same as the previous figure, while the horizontal dashed line represents the value $\langle \kappa_\Delta \rangle = \langle \kappa \rangle$. The different networks correspond to (a) A hospital (LH10), (b) a workplace in 2015 (InVS15), (c) a primary school (Lyon school), (d) a scientific conference (SFHH), (e) a workplace in 2013 (InVS13), and a (f) high-school in 2013 (Thiers13).

Figure 8 considers two new cases of social contacts: the quantitative assessment of contact patterns in a village in rural Malawi [6] and face-to-face contacts collected in an office building [7]. Note that, in both cases, $\langle \kappa_\Delta \rangle - \langle \kappa \rangle$ shows negative values before the percolation threshold of the network, indicating the possibility of reaching a second-order phase transition in the system.

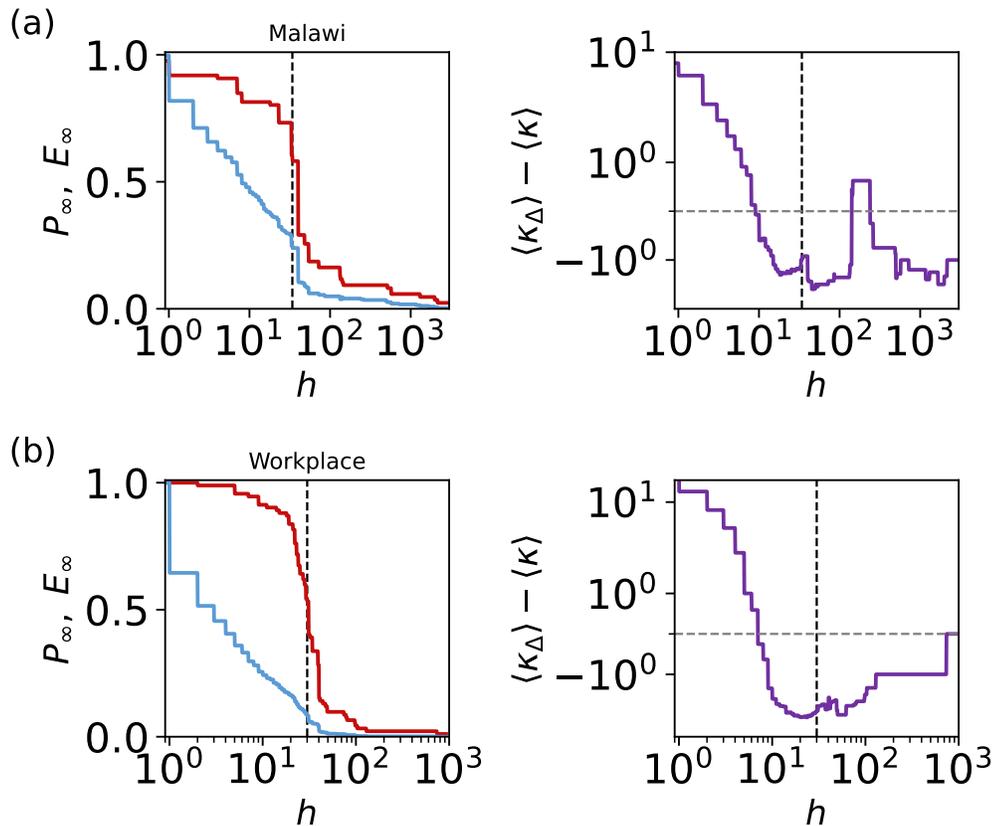

FIG. 8. **Real networks.** (Left) Giant cluster of the network ($P_\infty$) (red line) and relative number of edges, $E_\infty$, (blue line) versus threshold weight $h$. The black dashed line shows the value $h$ for which the giant component of the network has only 50% of the original number of individuals. (Right) $\langle \kappa_\Delta \rangle - \langle \kappa \rangle$ (violet line) versus threshold weight $h$. The black dashed line represents the same as the previous figure, while the horizontal dashed line represents the value $\langle \kappa_\Delta \rangle = \langle \kappa \rangle$. The different networks correspond to (a) a village in rural Malawi and (b) a workplace.

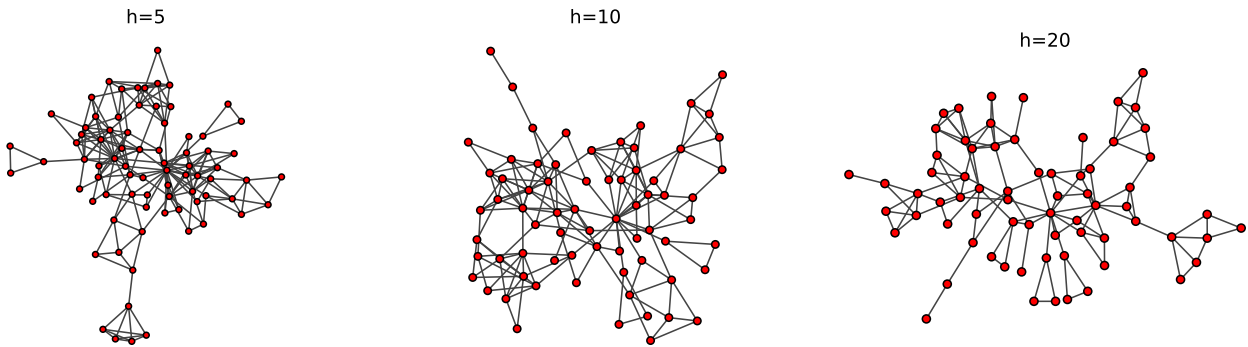

FIG. 9. **Contact network of a rural village in Malawi.** Sparsified network for different values of $h$: (a) 5, (b) 10, and (c) 20.

Figure 10 shows the results of face-to-face interactions collected on Thursday, October 1st, and Friday, October 2nd, 2009, of 77,602 contact events between 242 individuals (232 children and ten teachers) in a high school [8]. As shown in Figure 10, again $\langle\kappa_\Delta\rangle - \langle\kappa\rangle$ shows negative values before the percolation threshold of the network, indicating the possibility of reaching a second-order phase transition in the system.

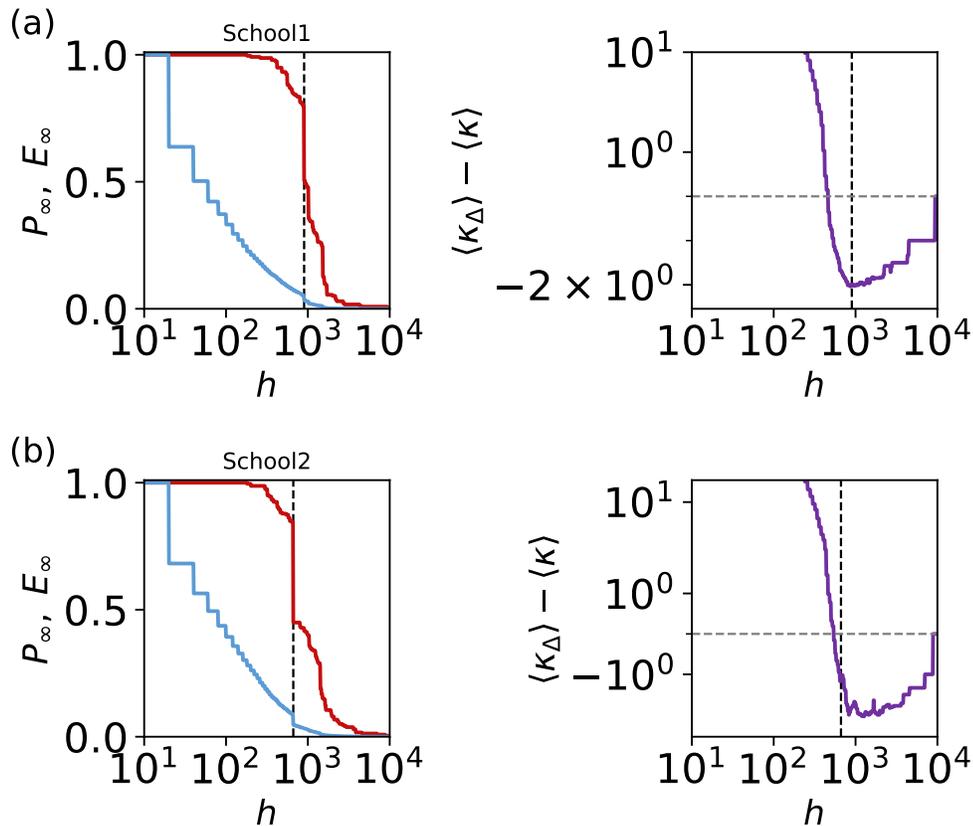

FIG. 10. **Real networks.** (Left) Giant cluster of the network ($P_\infty$) (red line) and relative number of edges, $E_\infty$, (blue line) versus threshold weight $h$. The black dashed line shows the value $h$ for which the giant component of the network has only 50% of the original number of individuals. (Right) $\langle\kappa_\Delta\rangle - \langle\kappa\rangle$ (violet line) versus threshold weight $h$. The black dashed line represents the same as the previous figure, while the horizontal dashed line represents the value $\langle\kappa_\Delta\rangle = \langle\kappa\rangle$. The different networks correspond to a high school on two consecutive days: (a) Thursday, October 1st, and (b) Friday, October 2nd, 2009.

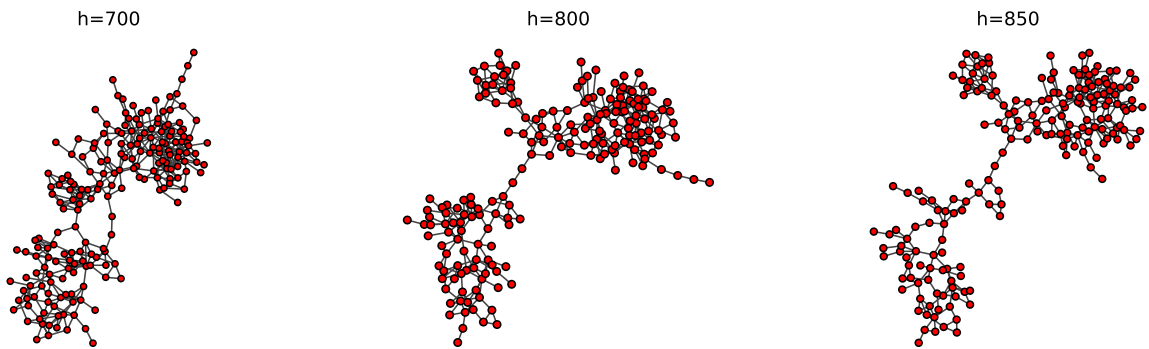

FIG. 11. **Contact network of a high school in France.** Sparsified network for School1 and different values of $h$: (a) 700, (b) 800, and (c) 850.

Figure 12 shows the results of face-to-face interactions collected between high school students of several classes in Lycée Thiers, Marseilles, France, during four days (Tuesday to Friday) in Dec. 2011 and during seven days (from a Monday to the Tuesday of the following week) in Nov. 2012 [9]. As shown in Figure 12, $\langle \kappa_\Delta \rangle - \langle \kappa \rangle$ shows negative values before the network percolation threshold, indicating the possibility of reaching a second-order phase transition in the system.

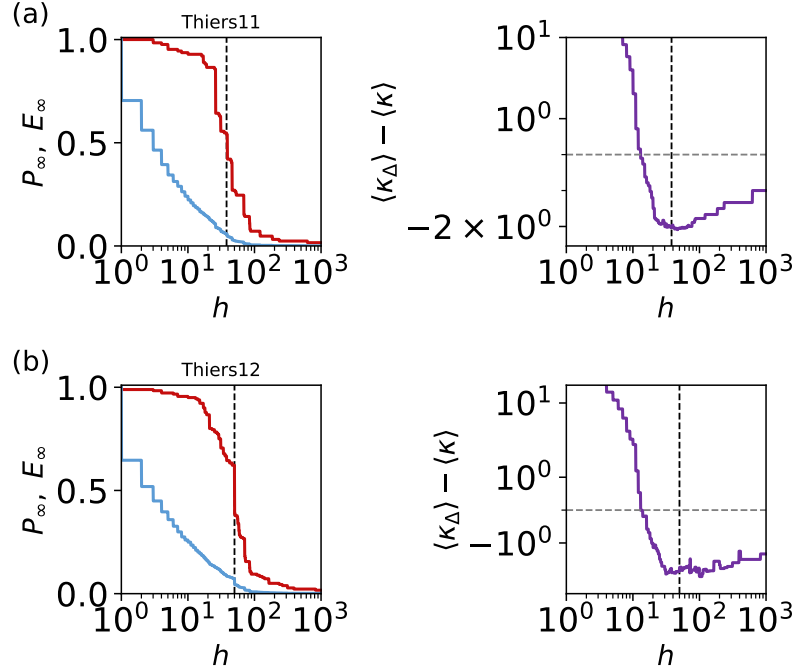

FIG. 12. **Real networks.** (Left) Giant cluster of the network ($P_\infty$) (red line) and relative number of edges, $E_\infty$, (blue line) versus threshold weight $h$. The black dashed line shows the value $h$ for which the giant component of the network has only 50% of the original number of individuals. (Right) $\langle \kappa_\Delta \rangle - \langle \kappa \rangle$ (violet line) versus threshold weight $h$. The black dashed line represents the same value as the previous figure, while the horizontal dashed line represents the $\langle \kappa_\Delta \rangle = \langle \kappa \rangle$. The different networks correspond to a high school for two consecutive years: (a) 2011 and (b) 2012.

The last data set analyzed involves high school students in specific classes called "classes préparatoires" in Lycée Thiers, Marseilles, France [10]. As shown in Figure 13, $\langle \kappa_\Delta \rangle - \langle \kappa \rangle$ shows negative values before the network percolation threshold, indicating the possibility of reaching a second-order phase transition in the system.

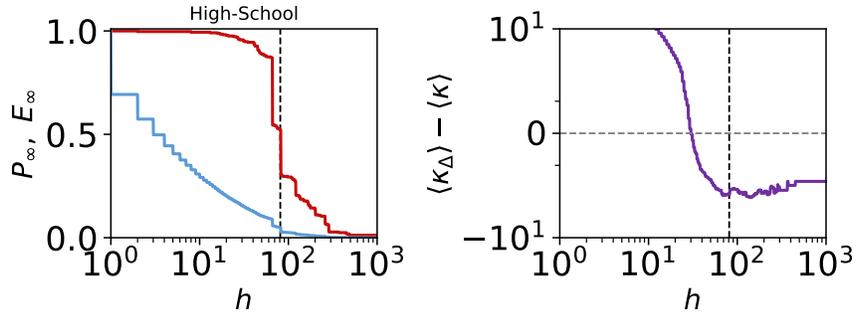

FIG. 13. **Real networks.** (Left) Giant cluster of the network ($P_\infty$) (red line) and relative number of edges, $E_\infty$, (blue line) versus threshold weight $h$. The black dashed line shows the value $h$ for which the giant component of the network has only 50% of the original number of individuals. (Right) $\langle \kappa_\Delta \rangle - \langle \kappa \rangle$ (violet line) versus threshold weight $h$. The black dashed line represents the same value as the previous figure, while the horizontal dashed line represents the $\langle \kappa_\Delta \rangle = \langle \kappa \rangle$. The different networks correspond to a high-school Lycée Thiers.

## VII. HYPEREDGE OVERLAP FOR DIFFERENT SYNTHETIC NETWORKS

Here, we have analyzed the recently introduced Hyperedge overlap $T_i^{(m)}$, a measure accounting for the overlap of nodes that belong to the same hyperedge [11]. Figure 14 shows the averaged Hyperedge overlap over nodes, $\langle T \rangle$, for different networks such as BA and HMN. As observed, for BA networks $\langle T \rangle$ presents a continuous decay compatible with the vanishing clustering coefficient. In contrast, networks with a low spectral dimension, as in the case of HMN, present high values of $\langle T \rangle$ even if the network still presents a second-order phase transition in the infinite-size limit (see Supp. Figs. 5 and 6).

In particular, for the case of BA networks, Figure 14 cannot predict the change in the order of the phase transition due to finite-size effects, as explicitly stated in Figure 2 of the main text when analyzing the change of sign in $b$. At the same time, Fig. 14(b) and (c) demonstrate a conflicting result for the case of HMN networks. That is, for different values of $m_0$, the hyperedge overlap increases ($m_0 = 3$) or decreases ($m_0 = 4$) depending on $\alpha$, even if, in both cases (as previously shown in Suppl. Figs. 5 and 6) there exists a specific value when $\alpha$ increases that generates the emergence of the first-order phase transition. This again shows that for low-dimensional systems, field-theoretic descriptions are those capturing the possible emergent behavior of the system.

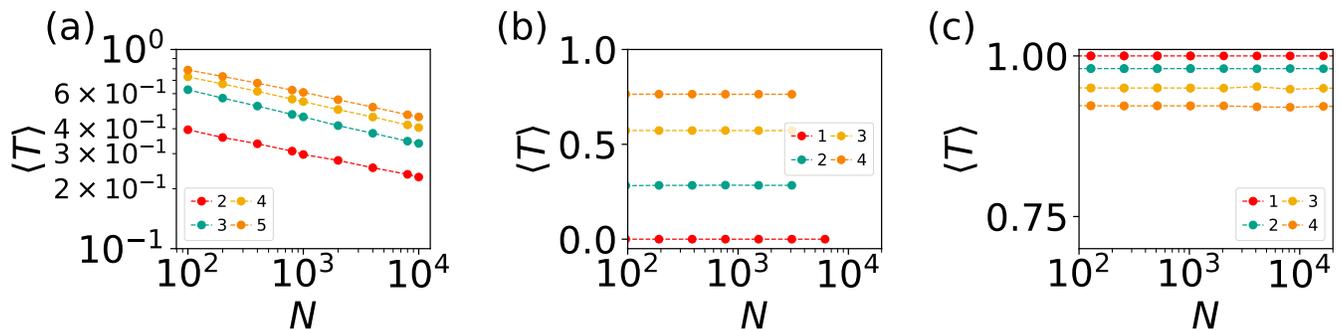

FIG. 14. **Hyperedge overlap.** Mean hyperedge overlap versus system size for (a) BA networks with different values of m (see legend), (b) HMN networks with $M_0 = 3$ and different values of $\alpha$, and (c) HMN networks with $M_0 = 4$ and different values of $\alpha$. Note that for all cases, it is not predictive of the type of phase transition in the system. All curves have been averaged over $10^3$ independent realizations.

- [1] I. Iacopini, G. Petri, A. Barrat, and V. Latora, Simplicial models of social contagion, Nat. Comm. **10**, 2485 (2019).
- [2] H. Hinrichsen, Non-equilibrium critical phenomena and phase transitions into absorbing states, Adv. Phys. **49**, 815 (2000).
- [3] S. Di Santo, P. Villegas, R. Burioni, and M. A. Muñoz, Landau–ginzburg theory of cortex dynamics: Scale-free avalanches emerge at the edge of synchronization, Proc. Natl. Acad. Sci. U.S.A. **115**, E1356 (2018).
- [4] P. Moretti and M. A. Muñoz, Griffiths phases and the stretching of criticality in brain networks, Nat. Commun. **4**, 2521 (2013).
- [5] M. Génois and A. Barrat, Can co-location be used as a proxy for face-to-face contacts?, EPJ Data Sci. **7**, 1 (2018).
- [6] L. Ozella, D. Paolotti, G. Lichand, J. P. Rodríguez, S. Haenni, J. Phuka, O. B. Leal-Neto, and C. Cattuto, Using wearable proximity sensors to characterize social contact patterns in a village of rural malawi, EPJ Data Sci. **10**, 46 (2021).
- [7] M. Génois, C. L. Vestergaard, J. Fournet, A. Panisson, I. Bonmarin, and A. Barrat, Data on face-to-face contacts in an office building suggest a low-cost vaccination strategy based on community linkers, Netw. Sci. **3**, 326 (2015).
- [8] J. Stehlé, N. Voirin, A. Barrat, C. Cattuto, L. Isella, J.-F. Pinton, M. Quaggiotto, W. Van den Broeck, C. Régis, B. Lina, *et al.*, High-resolution measurements of face-to-face contact patterns in a primary school, PloS ONE **6**, e23176 (2011).
- [9] J. Fournet and A. Barrat, Contact patterns among high school students, PloS ONE **9**, e107878 (2014).
- [10] R. Mastrandrea, J. Fournet, and A. Barrat, Contact patterns in a high school: a comparison between data collected using wearable sensors, contact diaries and friendship surveys, PloS ONE **10**, e0136497 (2015).
- [11] F. Malizia, S. Lamata-Otín, M. Frasca, *et al.*, Hyperedge overlap drives explosive transitions in systems with higher-order interactions, Nat. Commun. **16**, 555 (2025).



\end{document}